\documentclass[11pt, a4paper]{article}
\usepackage[comma,authoryear]{natbib}
\usepackage{amsmath, amssymb, amsfonts, verbatim, graphicx,
  rotating, multirow, a4}

\newtheorem{lem}{Lemma}[section]
   {\begin{trivlist}\item[]{ series\sffamily Keywords:}\ }
   {\end{trivlist}}
\usepackage{xr}
\begin{document}

\title{  Using The Censored Gamma Distribution for Modeling Fractional Response Variables with
  an Application to Loss Given Default  }

\author{Fabio Sigrist,  Werner A. Stahel\\ Seminar for Statistics, ETH Z\"urich}
\date{}
\maketitle
 
\begin{abstract}
Regression models for limited continuous dependent variables having a
non-negligible probability of attaining exactly their limits are
presented. The models differ in the number of parameters and in their
flexibility.    Fractional data being a
special case of limited dependent data, the models also apply to variables
that are a fraction or a proportion.   It is shown how to fit these models and they are applied to a Loss
Given Default dataset from insurance to which they provide a good fit.
\end{abstract}

\noindent\textsc{Keywords}: {Fractional response variables; censored distributions; Tobit models; limited dependent variables; Loss Given Default}


\section{Introduction}
 
Proportions or fractions are of considerable interest in economics as well
as other sciences. They are usually bounded by 0 and 1 (or 100\%).
Often, such quantities show a substantial probability for adopting one or
both of the boundary values. Such variables have been termed
``fractional response variables'' by \citet{PaWo96}. In a recent survey paper on
modeling fractional data, \citet{RaRaMu11} list pension plan participation
rates, firm market share, proportion of debt in the financing mix of firms,
fraction of land area allocated to agriculture, and proportion of exports
in total sales as examples. Another example is illustrated in
\citet{PaWo08}, where test pass rates are analyzed.

In insurance, losses are frequently restricted to be positive and below an
upper bound defined by a contract. We analyze a
Loss Given Default dataset from an insurance category called ``surety''. In
this example, claims cannot exceed a prespecified insured maximum, i.e., the ratio of loss over maximum is bounded by
1. On the other hand, for several reasons, the claims often do not
lead to ultimate losses. The interest is in
relating the distribution of this variable to a set of explanatory variables by a regression model.

For a fractional response variable $Y$, an important type of models focuses on
the conditional mean $E[Y|\boldsymbol{x}]$ given a vector of covariates $\boldsymbol{x}$. A popular choice is to
use the logistic function as a link function between a linear predictor and $E[Y|\boldsymbol{x}]$, but other cumulative distribution functions
can also be used. Another semiparametric approach relies on assumptions about quantiles (see, e.g., \citet{Po84}, \citet{KhPo01} or \citet{ChKh01}). Whereas these approaches are sufficient for the purpose of
many studies,
in other cases, other aspects of the distribution of $Y$ given $\boldsymbol{x}$, like
upper quantiles or probabilities of attaining the limits, are of
interest, as is the case in our application. In that case, parametric
models are advantageous. On the other hand, since semiparametric models rely on
less assumptions, they have the advantage that they are less prone to misspecification.


When there is a non-zero probability that the boundary
values are attained, it is natural to use models based on censored random
variables. These models are used in different fields of application. In economics, analyzing household expenditure
on durable goods, \citet{To58} first introduced such a
model which later was coined Tobit model by \citet{Go64}. In climate
science, precipitation can be modelled using censored distributions (see,
e.g., \citet{BaPl92} or \citet{SaGu04}). 

The Tobit model describes the distribution of $Y$ given $\boldsymbol{x}$
as a censored normal with expectation $\mu=\boldsymbol{x}'\boldsymbol{\beta}$.
It is therefore often perceived as a model for censored data, which it is
in the detection limit case.  However, it is perfectly adequate to
use the censored normal distribution as a probability model in situations
where no actual censoring occurs and the zeros are genuine values of the
response, as is the case for the original application of \citet{To58}. The use of censored distributions is then a device to
obtain a tractable model even though the data is not actually censored.

 To support our thinking about the situation to be modeled, it
is often helpful to attach the idea of a ``potential'' to a latent,
uncensored response variable $Y^*$, of which $Y$ is the censored version. In the case of precipitation, this potential
measures a tendency for rain which may move from zero to way below,
indicating that the weather develops from cloudy to very dry. For the standardized losses
in our insurance example, the latent variables can be thought of as a loss
potential.  We note that \citet{Wo10} calls models for variables that have a
discrete and a continuous part, without actual censoring occurring, corner solution
models. In \citet[Chapter 16]{Wo02}, it is stated that an additional
advantage of using a parametric distribution for modeling corner solution outcomes is that estimates of
quantities such as $E[Y|\boldsymbol{x}]$ are efficient.

The Tobit model is easily adjusted to the case of an additional upper limit
for $Y$ (\citet{RoNe75}) and thus to fractional data, and the generalization to
replacing the normal distribution by any other suitable family is
conceptually straightforward. In fact, when using
censored distributions, one can model all quantities of interest, such as
the mean, quantiles, and probabilities of attaining limits, together.
We will focus on this approach in the following, using a shifted gamma
distribution instead of the normal.
The gamma distribution is a flexible distribution that is popular in
insurance, and it will be shown to fit our data well (see, e.g., Figure
\ref{LGDvsFVfit}).


Models based on a single random variable, such as the Tobit model or the censored gamma model, have the advantage of having a parsimonious parameterization
entailing easier and more consistent interpretation. However, there are situations in which the
frequencies of the limits do not follow this parsimonious description. We therefore also introduce two extensions of the model. For instance, in our example there may even be administrative reasons
for an excessive number of zero losses, due to incentives to place a claim with little justification. Such preventive filing may result in a large
number of ``additional zeros''. This idea suggests a mixture model,
consisting of a censored part, as introduced above, and a model for the
additional zeros.

An other approach to tackle this problem is called two-part models by
\citet{RaRaMu11}. These are extensions of the models of \citet{PaWo96}. Here, a first model describes the occurrence of boundary
values. Then, the continuous part can be modeled, for instance, by using
the beta distribution (see \citet{Pa01} and \citet{FeCr04}). \citet{RaSi09}
and \citet{CoKiMc08} present empirical applications of two-part fractional
response models. We also introduce an alternative extension of the censored
gamma model based on this two-part modeling idea. Here, the probabilities of attaining the
boundary value(s) are modelled separately from the continuous part in
between them.

The rest of the paper is organized as follows. In Section \ref{SecCGMod}, we introduce the censored gamma model, show how it can be
interpreted, and derive an estimation procedure for it. In Section
\ref{SecExtCGMod}, two possible generalizations are
presented. In Section \ref{SecAppl}, we illustrate an application of the
models to the dataset mentioned above.


\section{The Censored Gamma Model}\label{SecCGMod}
In order to establish ideas, consider the Tobit model in its two sided version as developed by \citet{RoNe75}. It is assumed that there exists a latent
variable  $Y^*$ which is, conditional on some covariates $\boldsymbol{x}=(x_1,\dots,x_p) \in \mathbb{R}^p$,
normally distributed. This variable is observed only
if it lies in the interval $[0,1]$. Otherwise, we observe $0$ or
$1$, depending on whether the latent variable is smaller than $0$  or
greater than $1$, respectively. If $Y$ denotes the observed variable, this can be
expressed as
\begin{equation}\label{latenteq}
Y^*|\boldsymbol{x} \sim \mathcal{N}(\mu,\sigma^2)
\end{equation}
and
\begin{equation}\label{censeq}
\begin{split}
  Y&=0,~~ \text{if}~Y^* \leq 0,\\
  &=Y^*,~~ \text{if}~ 0 < Y^* < 1,\\
  &=1,~~ \text{if}~Y^* \geq 1.
\end{split}
\end{equation}

Furthermore, the expectation $\mu$ of the latent variable $Y^*$ is related to the covariates
$\boldsymbol{x}$ through
$$\mu=\boldsymbol{x}'\boldsymbol{\beta},~~\boldsymbol{\beta}\in\mathbb{R}^p.$$ 

For more details, e.g., on
inference, we refer to \citet{Ma83}, Chapter 6, and \citet{Am85},
Chapter 10.  Furthermore, \citet{Br96} and \citet{Lo97} give
overviews of models for limited dependent variables.

Clearly, the assumption of a normal distribution for $Y^*$ is not adequate
for all data. It is well known that the Tobit model is sensitive to
distributional assumptions (see, e.g., \citet {ArSc82} or
\citet{MaNe75}). A natural   alternative   is to replace the normal
distribution by another one. We choose a shifted gamma distribution since
it is a flexible distribution that is applied in many areas, especially in
insurance. Further, it provides a good
fit to the dataset of insurance claims mentioned above.   This
choice relies on distributional assumptions which have to be checked when
applying the model to data.    

To avoid unnecessary inflation of notation, we let the boundaries of the observed
variable be $0$ and $1$. The model is easily generalized for variables whose
range of values is any interval $[y_l,y_u]$ with $y_l<y_u$, though. This might be done
either by first applying a linear transformation to the respective variable or by
reformulating the model. The case where the observations are only bounded
from below is included by letting $y_u\rightarrow \infty$. 

\subsection{The Model}
Generalizing the Tobit model specified in \eqref{latenteq} and \eqref{censeq}, it is assumed that there exists a latent variable $Y^*$ which has,
conditional on $\boldsymbol{x}$, a distribution with density
$f^*_{\boldsymbol{\theta}^*}(y^*)$ and cumulative distribution function
$F^*_{\boldsymbol{\theta}^*}(y^*)$, $\boldsymbol{\theta}^*$ being a
vector of parameters. The observed dependent variable $Y$ then depends on the
latent variable as in \eqref{censeq}.

It follows that the distribution of such a censored variable $Y$ can be
characterized by
\begin{equation}\label{CGprob}
\begin{split}
P[Y=0]&=F^*_{\boldsymbol{\theta}^*}(0),\\
P[Y\in (y,y+dy)]&=f^*_{\boldsymbol{\theta}^*}(y)dy,~~0<y<1,\\
P[Y=1]&=1-F^*_{\boldsymbol{\theta}^*}(1).
\end{split}
\end{equation}

Consequently, the density of the observed variable $Y$ can be written as
\begin{equation}\label{CGdens1}
\begin{split}
f_{\boldsymbol{\theta}^*}(y)=&F^*_{\boldsymbol{\theta}^*}(0)\delta_0(y)+f^*_{\boldsymbol{\theta}^*}(y)\mathbf{1}_{\{0<
  y<1\}}(y)+(1-F^*_{\boldsymbol{\theta}^*}(1))\delta_1(y), ~~0\leq y \leq 1,
\end{split}
\end{equation}
where $\delta_0(y)$ and $\delta_1(y)$ are Dirac measures and where
$\mathbf{1}_{\{0< y<1\}}(y)$ denotes the indicator function equaling $1$ if
$0<y<1$ and $0$ otherwise.

In order to extend the model to the regression case, we relate the
distribution of $Y^*$ to the
covariates $\boldsymbol{x}$. This is done by assuming
that the main parameter $\vartheta$ of the distribution of $Y^*$, which
might be the mean
or a scale parameter, is related through a link function $g$ to the
covariates,
\begin{equation}
g(\vartheta)=\boldsymbol{x}'\boldsymbol{\beta}.
\end{equation}

In the following, we will focus on the
case where the distribution of $Y^*$ is specified as a gamma distribution
with a shifted origin. The density and the distribution function of a gamma distributed variable
with shape parameter $\alpha$ and scale parameter $\vartheta$ are denoted by
$g_{\alpha,\vartheta}(y)$ and $G_{\alpha,\vartheta}(y)$, respectively. The
density of a shifted gamma distribution is then
$$g_{\alpha,\vartheta}(y^*+\xi)=\frac{1}{\vartheta^{\alpha}\Gamma(\alpha)}(y^*+\xi)^{\alpha-1}e^{-(y^*+\xi)/\vartheta},~~y^*>-\xi,$$
where $\xi,\vartheta,\alpha>0$, and its distribution function is $G_{\alpha,\vartheta}(y^*+\xi)$.

The density of the observed $Y$ can be expressed as
\begin{equation}\label{CGdens2}
\begin{split}
f_{\alpha,\vartheta,\xi}(y)=  & G_{\alpha,\vartheta}(\xi)\delta_0(y)+g_{\alpha,\vartheta}(y+\xi)\mathbf{1}_{\{0<
  y<1\}}(y)\\ &+(1-G_{\alpha,\vartheta}(1+\xi))\delta_1(y), ~~0\leq y \leq 1.
\end{split}
\end{equation}

The use of a gamma distribution with a shifted origin,
instead of a standard gamma distribution, is motivated by the
fact that the lower censoring occurs at zero. In this case, the shift
$\xi$ is needed to obtain a positive probability of $Y=0$.

For the regression case, we assume that the scale parameter $\vartheta$ is related to the
covariates via the logarithmic link function
\begin{equation}
\log(\vartheta)=\boldsymbol{x}'\boldsymbol{\beta}.
\end{equation}

Henceforth and if not otherwise stated, we assume that $Y^*$ (and $Y$) follow a
(censored) shifted
gamma distribution. We will refer to this model as the ``censored gamma model''.

Note that if no censoring occurred and $\xi$ was set to zero, the
censored gamma model
would be a generalized linear model (\citet{McNe83}) for a gamma distributed
variable with a logarithmic link function.

\subsection{Interpretation}\label{CGInterpr}
If the focus lies on the latent response variable
$Y^*$, the interpretation is straightforward. Since
\begin{equation}\label{expG}
E[Y^*|\boldsymbol{x}]=\alpha\vartheta - \xi,
\end{equation}
the marginal effect of a continuous predictor $x_j$ on  
$E[Y^*|\boldsymbol{x}]$   is
\begin{equation}\label{marG}
\frac{\partial E[Y^*|\boldsymbol{x}]}{\partial x_j}=\beta_j \alpha\vartheta .
\end{equation}

On the other hand, one might be primarily interested
in the observed variable $Y$, rather than the latent variable $Y^*$. Its
mean and corresponding marginal effects are calculated in the following lemma.

\begin{lem}\label{LemIntp}
The following holds true.
\begin{equation}\label{expCG} 
\begin{split}
E[Y|\boldsymbol{x}]=&\alpha\vartheta\left(G_{\alpha+1,\vartheta}(1+\xi)-G_{\alpha+1,\vartheta}(\xi)\right)\\
&+(1+\xi)\left(1-G_{\alpha,\vartheta}(1+\xi)\right)-\xi\left(1-G_{\alpha,\vartheta}(\xi)\right),
\end{split}
\end{equation}
and for a continuous covariate $x_j$,
\begin{equation}\label{marCG}
\frac{\partial E[Y|\boldsymbol{x}]}{\partial x_j}=\beta_j\alpha\vartheta(G_{\alpha+1,\vartheta}(1+\xi)-G_{\alpha+1,\vartheta}(\xi)).
\end{equation}
\end{lem}
The derivation of these two equations is shown in Appendix \ref{ProofLem}.

We note that the marginal effect of $x_j$ on $E[Y|\boldsymbol{x}]$ is a scaled version
of the effect on $E[Y^*|\boldsymbol{x}]$, with the scaling factor depending
nonlinearly on the covariates.

If the interest lies on, say, the probability of $Y$ being zero,
$P[Y=0]=G_{\alpha,\vartheta}(\xi)$, one can also calculate partial effects
on this quantity. For a continuous $x_j$, using similar ideas as in the proof of
the above lemma, it is easily shown that
\begin{equation}
\begin{split}
\frac{\partial P[Y=0|\boldsymbol{x}]}{\partial x_j}&=\frac{\partial
  G_{\alpha,\vartheta}(\xi)}{\partial x_j}\\
&=-\beta_j \xi g_{\alpha,\vartheta}\left(\xi\right).
\end{split}
\end{equation}

Finally, one can also consider quantiles. The quantile function
$F^{\leftarrow}_{\alpha,\vartheta,\xi}(q)$, for $q\in [0,1]$, of $Y$ is given by
\begin{equation}
\begin{split}
  F^{\leftarrow}_{\alpha,\vartheta,\xi}(q)&=0,~~ ~~ ~~  ~~ ~~ ~~  ~~ ~
  \text{if}~0\leq q \leq G_{\alpha,\vartheta}(\xi),\\
  &=\vartheta G^{-1}_{\alpha,1}(q)-\xi,~~ \text{if}~  G_{\alpha,\vartheta}(\xi) < q <  G_{\alpha,\vartheta}(1+\xi),\\
  &=1,~~ ~~ ~~ ~~  ~~ ~\text{if} ~~   G_{\alpha,\vartheta}(1+\xi) \leq
  q \leq 1.
\end{split}
\end{equation}
The partial effect of a continuous covariate $x_j$ on the $q$-quantile
$F^{\leftarrow}_{\alpha,\vartheta,\xi}(q)$ is therefore
\begin{equation}
\begin{split}
  \frac{\partial F^{\leftarrow}_{\alpha,\vartheta,\xi}(q)}{\partial x_j}&=0,~~ ~~  ~~ ~~ ~~  ~~ ~
  \text{if}~0< q < G_{\alpha,\vartheta}(\xi),\\
  &=\beta_j \vartheta
G^{-1}_{\alpha,1}(q),~~ \text{if}~  G_{\alpha,\vartheta}(\xi) < q <  G_{\alpha,\vartheta}(1+\xi),\\
  &=0,~~ ~~ ~~ ~~  ~~ ~~~\text{if} ~~   G_{\alpha,\vartheta}(1+\xi) <
  q < 1.
\end{split}
\end{equation}

Note that for the cases $q = G_{\alpha,\vartheta}(\xi)$ and $q =
G_{\alpha,\vartheta}(1+\xi)$, the function
$F^{\leftarrow}_{\alpha,\vartheta,\xi}(q)$ is not differentiable with
respect to $x_j$ and, consequently, partial effects cannot be calculated.
 

\subsection{Estimation}
In this section, it is shown how to perform maximum likelihood estimation
for the censored gamma model using a Newton-Raphson method known as Fisher's scoring algorithm (see, e.g., \citet{FaTu01}).

Denoting generically by $\boldsymbol{\theta}$ all parameters that are to be
estimated and by $\ell(\boldsymbol{\theta})$ the log-likelihood,
Fisher's scoring algorithm starts with an initial estimate
$\hat{\boldsymbol{\theta}}^{(0)}$ and iteratively calculates (until
convergence is achieved)
$$\hat{\boldsymbol{\theta}}^{(k+1)}=\hat{\boldsymbol{\theta}}^{(k)}+I\left(\hat{\boldsymbol{\theta}}^{(k)}\right)^{-1}s\left(\hat{\boldsymbol{\theta}}^{(k)}\right),~~k=0,1,2,\dots,$$
where $$s\left(\boldsymbol{\theta}\right)=\frac{\partial
  \ell(\boldsymbol{\theta})}{\partial \boldsymbol{\theta}}$$ denotes the
score function, i.e., the first derivative of the log-likelihood, and
$$I\left(\boldsymbol{\theta}\right)=E_{\boldsymbol{\theta}}\left[s\left(\boldsymbol{\theta}\right)s\left(\boldsymbol{\theta}\right)^T \right]$$ is
the Fisher Information Matrix. How these two quantities are calculated for
the censored gamma model is shown in the following.

First, we reparametrize the shape parameter $\alpha$ through
\begin{equation}
\alpha'=\log(\alpha)
\end{equation}
to ensure that $\alpha$ attains only positive values. The parameters that
are to be estimated, therefore, consist of
$\boldsymbol{\theta}=(\alpha',\boldsymbol{\beta},\xi)$.

Assuming that we have independent data $y_1,\dots,y_n$ with covariates
$\boldsymbol{x}_1,\dots,\boldsymbol{x}_n$, the
log-likelihood function can be written as
\begin{equation*}
\ell(\boldsymbol{\theta})=\sum_{i =1}^n{\ell_i(\boldsymbol{\theta})}.
\end{equation*}
\begin{lem}
The following relations hold true.
\begin{align}
\label{CGllDera}
\frac{\partial \ell_i(\boldsymbol{\theta})}{\partial
  \alpha'}= &\frac{\alpha}{G_{\alpha,\vartheta_i}(\xi)}\left(-\psi(\alpha)G_{\alpha,\vartheta_i}(\xi) +
  H^{(1)}_{\alpha}\left(0, \frac{\xi}{\vartheta_i}\right)\right)\mathbf{1}_{\{y_i=0\}}\\ \nonumber&+a
\left(-\log(\vartheta_i)-\psi(\alpha)+\log(y_i+\xi)\right)\mathbf{1}_{\{0<  y_i <  1\}}\\\nonumber &-\frac{\alpha}{1-G_{\alpha,\vartheta_i}(1+\xi)}\left(-\psi(\alpha)G_{\alpha,\vartheta_i}(1+\xi) +
  H^{(1)}_{\alpha}\left(0, \frac{1+\xi}{\vartheta_i}\right)\right)\mathbf{1}_{\{y_i=1\}},\\
\frac{\partial \ell_i(\boldsymbol{\theta})}{\partial \beta_k}
=&-x_{ik}\xi \frac{g_{\alpha,\vartheta_i}\left(\xi\right)}{
  G_{\alpha,\vartheta_i}(\xi)}\mathbf{1}_{\{y_i=0\}}+x_{ik}\left(-\alpha+\frac{y_i+\xi}{\vartheta_i}\right)\mathbf{1}_{\{0<  y_i <
  1\}}\label{CGllDerb}\\ &+x_{ik}(1+\xi)\frac{ g_{\alpha,\vartheta_i}\left(1+\xi\right)}{1-
  G_{\alpha,\vartheta_i}(1+\xi)}\mathbf{1}_{\{y_i=1\}},\nonumber\\
\label{CGllDerh}
\frac{\partial \ell_i(\boldsymbol{\theta})}{\partial \xi}&=\frac{g_{\alpha,\vartheta_i}(\xi)}{G_{\alpha,\vartheta_i}(\xi)}\mathbf{1}_{\{y_i=0\}}+\left(\frac{\alpha-1}{y_i+\xi}-\frac{1}{\vartheta_i}\right)\mathbf{1}_{\{0< y_i <  1\}}-\frac{g_{\alpha,\vartheta_i}(1+\xi)}{1-G_{\alpha,\vartheta_i}(1+\xi)}\mathbf{1}_{\{y_i=1\}},
\end{align}
where $$\psi(\alpha)=\frac{d \log(\Gamma(\alpha))}{d \alpha}$$ denotes the
digamma function (see \citet{AbSt64}) and the functions $H^{(1)}_{\alpha}$
and $H^{(2)}_{\alpha}$ are defined as\footnote{We note that the functions $H^{(1)}_{\alpha}(l,u)$ and $H^{(2)}_{\alpha}(l,u)$ can be
  calculated using numerical integration. In our application, we did this
  by adaptive quadrature using the QUADPACK routines 'dqags' and 'dqagi' (\citet{quadpack}) available from Netlib.}
\begin{equation}\label{Hdef}
H^{(1)}_{\alpha}(l,u):=\frac{1}{\Gamma(\alpha)}\int_l^u{\log(y)y^{\alpha-1}\exp(-y)dy}
\end{equation}
and
\begin{equation}\label{Idef}
H^{(2)}_{\alpha}(l,u):=\frac{1}{\Gamma(\alpha)}\int_l^u {\log(y)^2y^{\alpha-1}\exp(-y)dy}.
\end{equation}
\end{lem}
The derivation of the scoring functions is shown in the following.

At first, we infer from \eqref{CGprob} that the likelihood function of an
interval censored gamma distribution can be written as
\begin{equation}\
L_y(\alpha,\vartheta,\xi)=G_{\alpha,\vartheta}(\xi)\mathbf{1}_{\{y=0\}}+g_{\alpha,\vartheta}(y+\xi)\mathbf{1}_{\{0< y
  <1\}}+(1-G_{\alpha,\vartheta}(1+\xi))\mathbf{1}_{\{y=1\}}
\end{equation}
which is equivalent to writing
\begin{equation}\label{CGll}
L_y(\alpha,\vartheta,\xi)=G_{\alpha,\vartheta}(\xi)^{\mathbf{1}_{\{y=0\}}}\cdot
g_{\alpha,\vartheta}(y+\xi)^{\mathbf{1}_{\{0< y <1\}}}\cdot (1-G_{\alpha,\vartheta}(1+\xi))^{\mathbf{1}_{\{y=1\}}}.
\end{equation}
It follows that we can write the log-likelihood
function $\ell_i(\boldsymbol{\theta})$ of an observation $y_i$ as
\begin{equation*}
\begin{split}
\ell_i(\boldsymbol{\theta})=&\log(G_{\alpha,\vartheta_i}(\xi))\mathbf{1}_{\{y_i=0\}}+\log(g_{\alpha,\vartheta_i}(y_i+\xi))\mathbf{1}_{\{0<
  y_i <  1\}}\\ &+\log(1-G_{\alpha,\vartheta_i}(1+\xi))\mathbf{1}_{\{y_i=1\}}\\
=&\log(G_{\alpha,\vartheta_i}(\xi))\mathbf{1}_{\{y_i=0\}}\\ &+
\left(-\alpha\log(\vartheta_i)-\log(\Gamma(\alpha))+(\alpha-1)\log(y_i+\xi)-\frac{y_i+\xi}{\vartheta_i}\right)\mathbf{1}_{\{0<
  y_i <  1\}}\\
&+\log(1-G_{\alpha,\vartheta_i}(1+\xi))\mathbf{1}_{\{y_i=1\}},
\end{split}
\end{equation*}
where
$$\vartheta_i=\exp(\boldsymbol{x}_i'\boldsymbol{\beta})~~
\text{and}~~\alpha=\exp(\alpha').$$ 

The derivative of $\ell_i$ with respect to the
parameter $\alpha'$ in \eqref{CGllDera} is then calculated using the following identity.
\begin{align}
\frac{\partial G_{\alpha,\vartheta}(\xi)}{\partial \alpha}& = \frac{\partial
  G_{\alpha,1}\left(\frac{\xi}{\vartheta}\right)}{\partial \alpha}\nonumber\\
& = \frac{\partial }{\partial
  a}\left(\frac{1}{\Gamma(\alpha)}\int_0^{\xi/\vartheta}{y^{\alpha-1}\exp(-y)dy}\right)\nonumber\\
& = \frac{-\Gamma'(\alpha)}{\Gamma(\alpha)^2}\int_0^{\xi/\vartheta}{y^{\alpha-1}\exp(-y)dy}+\frac{1}{\Gamma(\alpha)}\int_0^{\xi/\vartheta}{\log(y)y^{\alpha-1}\exp(-y)dy}\nonumber\\
& = -\psi(\alpha)G_{\alpha,\vartheta}(\xi) + H^{(1)}_{\alpha}\left(0,\frac{\xi}{\vartheta}\right).\label{gammaDeriva}
\end{align} 
Next, using 
$$\frac{\partial \ell_i(\boldsymbol{\theta})}{\partial
    \beta_k}=\frac{\partial \ell_i(\boldsymbol{\theta})}{\partial
    \vartheta_i}\frac{\partial \vartheta_i}{\partial
    \beta_k}=\frac{\partial \ell_i(\boldsymbol{\theta})}{\partial
    \vartheta_i}\vartheta_ix_{ik}$$
and \eqref{gammaDerivs}, differentiating $\ell_i(\boldsymbol{\theta})$ with
respect to $\beta_k$ gives the result in \eqref{CGllDerb}. The
calculation of the derivative with respect to $\xi$ in \eqref{CGllDerh} is straightforward.

For the Fisher-scoring algorithm and for asymptotic inference, we calculate the Fisher
Information Matrix 
$$I(\boldsymbol{\theta})_{k,l}= E_{\theta}\left[ \frac{\partial \ell(\boldsymbol{\theta})}{\partial\theta_k}\frac{\partial \ell(\boldsymbol{\theta})}{\partial
    \theta_l} \right],~~1\leq k,l\leq 2+p.$$

Because of the independence of the observations, this can be written as
\begin{equation*}
\begin{split}
I(\boldsymbol{\theta})_{k,l}&=E_{\theta}\left[\left(\sum_{i
    =1}^n{\frac{\partial \ell_i(\boldsymbol{\theta})}{\partial\theta_k}}\right)\left(\sum_{i =1}^n{\frac{\partial \ell_i(\boldsymbol{\theta})}{\partial\theta_l}}\right)\right]\\ &=\sum_{i =1}^n{E_{\theta}\left[\frac{\partial \ell_i(\boldsymbol{\theta})}{\partial\theta_k}\frac{\partial \ell_i(\boldsymbol{\theta})}{\partial\theta_l}\right]}.
\end{split}
\end{equation*}

The specific calculations of the entries $E_{\theta}\left[\frac{\partial
    \ell_i(\boldsymbol{\theta})}{\partial\theta_k}\frac{\partial
    \ell_i(\boldsymbol{\theta})}{\partial\theta_l}\right]$ are shown in Section \ref{app1} in the supplementary material. 

As mentioned before, the Fisher Information Matrix $I(\boldsymbol{\theta})$
is used in the Fisher-scoring algorithm for fitting the model and for asymptotic inference, in particular to estimate
standard errors of the coefficients $\boldsymbol{\beta}$.

\section{Two Extensions of the Model}\label{SecExtCGMod}
A salient feature of the model defined in \eqref{CGprob}   and of the Tobit
model   is the assumption that the
same parameters govern both the behaviour of the uncensored values as well as
the probabilities of being censored from below or above. 

In order to relax this
assumption, various extensions have been proposed. Sample selection models, first introduced by \citet{He76}, are one
approach. \citet{CR71} came forward with another proposal relaxing the
aforementioned assumption of one set of parameters governing the entire
model. 

For count data, similar problems can arise: there may be more zeros than
expected by a simple model, which would otherwise fit well. Basically,
two different kinds of solutions have been put forward there. 

\citet{Ai55} first proposed to model the zeros and the
values bigger than zero separately. \citet{Mu86} used a mixture
consisting of a distribution for the whole range of data, including zeros,
and a point mass at zero to capture extra zeros. These two types of models have been extensively
applied in various areas of research including manufacturing defects (\citet{La92}),
patent applications (\citet{CrDu97}), road safety (\citet{Mi94}),
species abundance (\citet{WeCuDoLi96}), medical consultations
(\citet{Gu97}), use of recreational facilities (\citet{GuTr96};
\citet{ShSh96}), and sexual behaviour
(\citet{He94}). \citet{RiDeHi98} give an overview of these models.

Our two extensions are based on similar ideas. The main difference is the
way in which the zeros are modeled. In the first extension, the zeros and
the non-zero values are
modeled separately assuming that the mechanisms that govern the
probability of $Y$ being zero and the
non-zero part are different. In the second extension, the zeros are
modelled as a mixture of two mechanisms. One is responsible for artificial or extra zeros whereas the other part is the censored gamma model introduced
in Section \ref{SecCGMod}.

\subsection{The Two-tiered Gamma Model} 
Inspired by the approach of \citet{CR71}, we extend the model in \eqref{CGprob} by allowing for two different sets of parameters, one governing
the probability of $Y$ being zero, and the other the behaviour for $0<Y\leq 1$.

Alternatively, the model could also be extended by
allowing for a different set of parameters governing the probability of $Y$ being
one. The extension presented here, which we will call two-tiered gamma
model, is mainly motivated by the presumption that zeros are generated by
another mechanism than the one that governs the rest of the data. We remark
that the
extension to a ``three-tiered'' model including a different set of
parameters for governing the probability of $Y$ being one is straightforward.

More specifically, in the two-tiered gamma model, it is assumed that there
exist two latent variables 
$$Y^*_1 \sim G_{\alpha,\tilde{\vartheta}}(y^*_1+\xi),~~ \text{with}~~
\tilde{\vartheta}=\exp(\boldsymbol{x}'\boldsymbol{\gamma}),~~\boldsymbol{\gamma}\in \mathbb{R}^p$$
and
 $$Y^*_2 \sim
  G_{\alpha,\vartheta}(y^*_2+\xi)~\text{truncated at}~0,~~ \text{with} ~~\vartheta=\exp(\boldsymbol{x}'\boldsymbol{\beta}),~~\boldsymbol{\beta}\in \mathbb{R}^p.$$
The first latent variable $Y^*_1$ is again following a shifted gamma
distribution, whereas the second variable $Y^*_2$ has shifted gamma
distribution that is lower truncated at zero. These two latent variables are then related to $Y$ through
\begin{alignat*}{6}
Y&= 0 &\text{if}&  ~~Y^*_1 \leq &0,&\\
&= Y^*_2~~ &\text{if}&  ~~ &0&<Y^*_1 ~\text{and} ~Y^*_2<&1,&\\
&= 1 &\text{if}&  ~~&0&<Y^*_1~ \text{and} ~&1&\leq Y^*_2.
\end{alignat*}
In other words,  the two-tiered gamma model first decides whether $Y$ is zero or
not. This is modeled in the style of a probit model, using, however, a
cumulative gamma distribution function instead of a normal one. It is then assumed that, conditional on $Y>0$, $0<Y\leq 1$ has a lower truncated and upper
censored gamma distribution.

The distribution of $Y$ can then be characterized as follows.
\begin{equation}
\begin{split}
P[Y=0]=&G_{\alpha,\tilde{\vartheta}}(\xi),\\
P[Y\in
(y,y+dy)]=&g_{\alpha,\vartheta}(y+\xi)\frac{1-G_{\alpha,\tilde{\vartheta}}(\xi)}{1-G_{\alpha,\vartheta}(\xi)}dy,~~0<y<1,\\
P[Y=1]=&(1-G_{\alpha,\vartheta}(1+\xi))\frac{1-G_{\alpha,\tilde{\vartheta}}(\xi)}{1-G_{\alpha,\vartheta}(\xi)},
\end{split}
\end{equation}
with $$\vartheta=\exp(\boldsymbol{x}'\boldsymbol{\beta}),~ \tilde{\vartheta}=\exp(\boldsymbol{x}'\boldsymbol{\gamma}),~\boldsymbol{\beta},\boldsymbol{\gamma} \in \mathbb{R}^p,~\alpha,\xi>0.$$ 
Again, $g_{\alpha,\vartheta}(y)$ denotes the density of a Gamma$(\alpha,\vartheta)$ distributed variable and
$G_{\alpha,\vartheta}(y)$ is the corresponding distribution function. 

We remark that the distributions in both parts of the two-tiered model, i.e., the part modeling the
probability of $Y$ being zero and the part governing the behaviour of $0<Y
\leq 1$, are assumed to have the same shape parameter $\alpha$ and the same location
parameter $\xi$. Consequently, if $\boldsymbol{\beta}=\boldsymbol{\gamma}$, or $\vartheta= \tilde{\vartheta}$,
the two-tiered gamma model presented here and the aforementioned censored
gamma model coincide, which means that these two models are nested. This is
convenient for model comparison since it allows to use a likelihood
ratio test to compare the two models.

\subsection{Estimation of the Two-tiered Gamma Model}
 Having in mind that the censored gamma model is nested in the two-tiered
 gamma model,
we restrict ourselves to estimating the
coefficients $\boldsymbol{\beta}$ and $\boldsymbol{\gamma}$ of the two linear
predictors using Fisher's scoring algorithm. The shape parameter $\alpha$
and the location parameter $\xi$ could be estimated via numerical
optimization in an outer loop   with starting values obtained from first fitting
a censored gamma model.  

With $\boldsymbol{\theta}=(\boldsymbol{\beta},\boldsymbol{\gamma})$, the
log-likelihood function of the model can be written as
$\ell(\boldsymbol{\theta})=\sum_{i =1}^n{\ell_i(\boldsymbol{\theta})}$ with
\begin{equation*}
\begin{split}
\ell_i(\boldsymbol{\theta})=&\log(G_{\alpha,\tilde{\vartheta}_i}(\xi))\mathbf{1}_{\{y_i =
  0\}}\\ &+(\log(g_{\alpha,\vartheta_i}(y_i+\xi))+\log(1-G_{\alpha,\tilde{\vartheta}_i}(\xi))-\log(1-G_{\alpha,\vartheta_i}(\xi)))\mathbf{1}_{\{0< y_i <
  1\}}\\ &+(\log(1-G_{\alpha,\vartheta_i}(1+\xi)+\log(1-G_{\alpha,\tilde{\vartheta}_i}(\xi))-\log(1-G_{\alpha,\vartheta_i}(\xi)))\mathbf{1}_{\{y_i =
  1\}},
\end{split}
\end{equation*}
where
$$\vartheta_i=\exp(\boldsymbol{x}_i'\boldsymbol{\beta}),~~\tilde{\vartheta}_i=\exp(\boldsymbol{x}_i'\boldsymbol{\gamma}).$$
The score functions are
\begin{equation}
  \begin{split}
\frac{\partial \ell_i(\boldsymbol{\theta})}{\partial
  \beta_k}=&x_{ik}\left(\frac{y_i+\xi}{\vartheta_i}-a-\frac{\xi\cdot
    g_{\alpha,\vartheta_i}\left(\xi\right)}{1-
    G_{\alpha,\vartheta_i}(\xi)}\right)\mathbf{1}_{\{0<  y_i <  1\}}\\
&+x_{ik}\cdot\left(\frac{(1+\xi)\cdot  g_{\alpha,\vartheta_i}\left(1+\xi\right)}{1- G_{\alpha,\vartheta_i}(1+\xi)}-\frac{\xi\cdot  g_{\alpha,\vartheta_i}\left(\xi\right)}{1- G_{\alpha,\vartheta_i}(\xi)}\right)\mathbf{1}_{\{ y_i =  1\}}
\end{split}
\end{equation}
and
\begin{equation}
  \begin{split}
\frac{\partial \ell_i(\boldsymbol{\theta})}{\partial
  \gamma_k}=&-x_{ik}\frac{\xi\cdot
    g_{\alpha,\tilde{\vartheta}_i}\left(\xi\right)}{ G_{\alpha,\tilde{\vartheta}_i}(\xi)}\mathbf{1}_{\{ y_i =0\}}+x_{ik}\frac{\xi\cdot  g_{\alpha,\tilde{\vartheta}_i}\left(\xi\right)}{1- G_{\alpha,\tilde{\vartheta}_i}(\xi)}\cdot\left(\mathbf{1}_{\{0<  y_i <  1\}}+\mathbf{1}_{\{ y_i =1\}}\right).
\end{split}
\end{equation}
The entries of the Fisher
Information Matrix $I(\boldsymbol{\theta})$ are presented in Appendix \ref{app2}.

\subsection{The Zero-Inflated Gamma Model}
The extension presented in this section is motivated by the following idea. Assume that our
quantity of interest follows indeed a censored, shifted gamma
distribution. However, additional, artificial zeros occur by some other
mechanism and thus there are more zeros than expected.  \citet{DeIr84} used
such an extension of the Tobit model for modeling expenditures in household
budgets. $ $ Recently, a
zero-inflated model for censored continuous data has also been presented by
\citet{CoVi10}. 

These additional zeros are now allowed to follow their own model, in
contrast to the two-tiered model where all zeros were described together. This view may
make sense in specific applications like insurance, where some of the
claims that result in zero losses may be cases which were filed in order not
to miss a formal deadline or for similar artificial reasons.

In the zero-inflated model, the existence of two latent variables is again assumed,
$$Y^*_1 \sim N(-\mu,1)~~ \text{and} ~~ Y^*_2 \sim
  G_{\alpha,\vartheta}(y^*_2+\xi)$$ with 
  $\mu=\boldsymbol{x}'\boldsymbol{\gamma}$ and
  $\vartheta=\exp(\boldsymbol{x}'\boldsymbol{\beta})$. 

The censored
  gamma model is not nested in the zero-inflated model in the classical sense. However, the
  zero-inflated model coincides with the censored gamma model at the
  boundary of its parameter space, namely if $\mu \rightarrow -\infty$. For the reason of simplicity, we opt
  for the normal distribution. I.e., the extra zeros are model using a
  probit model. Alternatively,
  one could also use the logit distribution.

These two variables are then related to $Y$ through
\begin{alignat*}{8}
Y&= 0 &\text{if} & ~~Y^*_1 \leq &0,&~~~\text{or~ if}\\
& & & ~~&0&<Y^*_1~\text{and} ~Y^*_2\leq &0,&\\
&= Y^*_2 ~~&\text{if} & ~~&0&<Y^*_1~\text{and} ~&0&<Y^*_2<&1,&\\
&= 1 &\text{if} & ~~&0&<Y^*_1~\text{and}& & ~&1&\leq Y^*_2 .
\end{alignat*}

The variable $Y^*_1$ first decides whether the observed
response variable $Y$ is zero, i.e., if $Y^*_1\leq 0$ it follows that $Y=0$. Next, conditional on $Y^*_1>0$, $Y$ is distributed
according to a censored, shifted gamma distribution. 

This means that the
zeros are governed by two different components of the model. First, zeros can
arise if $Y^*_1$  is smaller than zero. And secondly, they can occur if, conditional on
$Y^*_1>0$, $Y^*_2$ is smaller than zero. Metaphorically speaking, we add extra mass at zero to the censored gamma
distribution, which can account for potential extra zeros. This approach
allows us to distinguish structural and extra zeros.

Note that the main distinctive feature of this model, in contrast to the
two-tiered model presented in the previous section, is that the distribution of the second
tier of the model is lower censored instead of lower truncated. 

As stated above, we choose to model the extra zeros using a probit model, i.e.,
\begin{equation}
p_0:=P[Y^*_1\leq 0]=\Phi(\boldsymbol{x}'\boldsymbol{\gamma}),~~\boldsymbol{\gamma} \in \mathbb{R}^p.
\end{equation}

Consequently, the distribution of $Y$ can be characterized by
\begin{equation}\label{ZIModel}
\begin{split}
P[Y=0]=&p_0+(1-p_0)\cdot G_{\alpha,\vartheta}(\xi),\\
P[Y\in (y,y+dy)]=&(1-p_0)\cdot
g_{\alpha,\vartheta}(y+\xi)dy,~~0<y<1,\\
P[Y=1]=&(1-p_0)\cdot(1-G_{\alpha,\vartheta}(1+\xi)),
\end{split}
\end{equation}
where $$p_0=\Phi(\boldsymbol{x}'\boldsymbol{\gamma}),~~\vartheta=\exp(\boldsymbol{x}'\boldsymbol{\beta}),~~\boldsymbol{\gamma},\boldsymbol{\beta}
\in \mathbb{R}^p,~~\alpha,\xi>0.$$

We note that the zero-inflated model reduces to the censored Gamma model in
the limit $\mu \to - \infty$, i.e., at the boundary of the parameter
space. This means that a straightforward likelihood ratio test for model
selection does not apply here. In Section \ref{ressec} in the application,
we use a simulation based testing procedure to compare these two models. 
 
\subsection{Estimation of the Zero-Inflated Gamma Model}
Since the EM (\citet{DeLaRu77}) algorithm lends itself naturally when it
comes to fitting mixtures of distributions and because calculations of
scores and the Fisher Information Matrix would be overly
complicated, we use the EM algorithm here.

The EM algorithms presented in the following finds the maximum likelihood estimators of the parameters
$\boldsymbol{\theta}=(\alpha,\boldsymbol{\beta},\boldsymbol{\gamma})$. The location parameter $\xi$
is fixed and assumed to be known. Again, $\xi$ could be obtained from first
fitting the censored gamma model or it could be estimated
through numerical optimization in an outer loop.   Alternatively, the values
obtained from the EM Algorithm together with the estimated $\xi$ from the censored
gamma model can be used as starting values for generic optimization
algorithms such as, for instance, quasi-Newton methods. We note that in some examples we observed convergence problems
when using quasi-Newton methods without reasonable starting values. 

With regard to the EM algorithm, we introduce two latent data
variables $Z$ and $Y^*$. For each $i$, $Z_i$ indicates whether the observation belongs to the extra
zero part of the model ($Z_i=0$) or to the censored gamma distribution
($Z_i=1$). The second missing data variable $Y^*_i$ is for the censored gamma part of
the model. It denotes the value of the underlying latent variable $Y^*_i$
which then is censored at zero and one. The complete data $\boldsymbol{W}$ therefore consists of
$(Z_1,Y^*_1),\dots,(Z_n,Y^*_n)$.

Using this, the complete-data likelihood can be written as
\begin{equation}
L_{\boldsymbol{W}}(\boldsymbol{\theta})=\prod_{i=1}^n{\left(\Phi(\boldsymbol{x}_i'\boldsymbol{\gamma})\right)^{1-Z_i}\cdot\left((1-\Phi(\boldsymbol{x}_i'\boldsymbol{\gamma}))\cdot
    g_{\alpha,\vartheta_i}(Y^*_i+\xi)\right)^{Z_i}},
\end{equation}
where $\log(\vartheta_i)=\boldsymbol{x}_i'\boldsymbol{\beta}$ and
$\boldsymbol{\theta}=(\alpha,\boldsymbol{\beta},\boldsymbol{\gamma})$, and
the complete-data log-likelihood is
\begin{equation}\label{ZIGEMll}
\begin{split}
\ell_{\boldsymbol{W}}(\boldsymbol{\theta})=&\sum_{i=1}^n{(1-Z_i)\log\left(\Phi(\boldsymbol{x}_i'\boldsymbol{\gamma})\right)+Z_i\log\left((1-\Phi(\boldsymbol{x}_i'\boldsymbol{\gamma}))
  \cdot g_{\alpha,\vartheta_i}(Y^*_i+\xi)\right)}\\
=&\sum_{i=1}^n{(1-Z_i)\log\left(\Phi(\boldsymbol{x}_i'\boldsymbol{\gamma})\right)+Z_i\log\left(1-\Phi(\boldsymbol{x}_i'\boldsymbol{\gamma})\right)}\\
&+\sum_{i=1}^n{Z_i\left(-\alpha\log(\vartheta_i)-\log(\Gamma(\alpha))+(\alpha-1)\log(Y^*_i+\xi)-\frac{Y^*_i+\xi}{\vartheta_i}\right)}.
\end{split}
\end{equation}

The EM algorithm produces a sequence of estimates
$\{\boldsymbol{\theta}^{(t)},~t=0,1,2,\dots\}$ by alternatively applying two
steps:

\textbf{E-step.} Compute the expected value of the log-likelihood, with respect to the conditional distribution of $\boldsymbol{W}$ given $\boldsymbol{y}$ under the current estimate of the parameters $\boldsymbol{\theta}^{(t)}$: $$Q^{(t+1)}(\boldsymbol{\theta})=E_{\boldsymbol{\theta}^{(t)}}\left[\ell_{\boldsymbol{W}}(\boldsymbol{\theta})|\boldsymbol{y}\right].$$
\textbf{M-step.} Update the parameter estimated according to:
$$\boldsymbol{\theta}^{(t+1)}=\text{argmax}_{\boldsymbol{\theta}} ~Q^{(t+1)}(\boldsymbol{\theta}).$$

From \eqref{ZIGEMll} , we infer that in the E-step three different expectations have to be calculated:
$E_{\boldsymbol{\theta}^{(t)}}\left[Z_i|\boldsymbol{y}\right]$,
$E_{\boldsymbol{\theta}^{(t)}}\left[Y^*_i+\xi|\boldsymbol{y}\right]$, and
$E_{\boldsymbol{\theta}^{(t)}}\left[\log(Y^*_i+\xi)|\boldsymbol{y}\right]$. For
the sake of notational brevity, we introduce the following two
abbreviations:

$$A_i^{(t)}=\Phi(\boldsymbol{x}_i'\boldsymbol{\gamma}^{(t)})$$
and 
$$B_i^{(t)}(\xi)=G_{\alpha^{(t)},\vartheta_i^{(t)}}(\xi).$$

The three expectations are then calculated as follows:
\begin{equation}
E_{\boldsymbol{\theta}^{(t)}}\left[Z_i|\boldsymbol{y}\right]=
\begin{cases}
\frac{(1-A_i^{(t)})\cdot
  B_i^{(t)}(\xi)}{A_i^{(t)}+(1-A_i^{(t)})\cdot
  B_i^{(t)}(\xi)} &\text{if}~~  y_i=0,\\
0 & \text{if}~~  y_i>0,\\
\end{cases}
\end{equation}

\begin{equation}
E_{\boldsymbol{\theta}^{(t)}}\left[Y^*_i+\xi|\boldsymbol{y}\right]=
\begin{cases}
\alpha^{(t)}\vartheta_i^{(t)}\frac{G_{\alpha^{(t)+1},\vartheta_i^{(t)}}(\xi)}{B_i^{(t)}(\xi)}&\text{if}~~  y_i=0,\\
y_i+\xi & \text{if}~~  0<y_i<1,\\
\alpha^{(t)}\vartheta_i^{(t)}\frac{1-G_{\alpha^{(t)+1},\vartheta_i^{(t)}}(1+\xi)}{1-B_i^{(t)}(1+\xi)}&\text{if}~~  y_i=1,\\
\end{cases}
\end{equation}
and
\begin{equation}
E_{\boldsymbol{\theta}^{(t)}}\left[\log(Y^*_i+\xi)|\boldsymbol{y}\right]=
\begin{cases}
 \log(\vartheta_i^{(t)})+\frac{H^{(1)}_{\alpha^{(t)}}\left(0, \frac{\xi}{\vartheta_i^{(t)}}\right)}{B_i^{(t)}(\xi)}&\text{if}~~  y_i=0,\\
\log(y_i+\xi) & \text{if}~~  0<y_i<1,\\
\log(\vartheta_i^{(t)})+\frac{H^{(1)}_{\alpha^{(t)}}\left(\frac{1+\xi}{\vartheta_i^{(t)}},\infty \right)}{1-B_i^{(t)}(1+\xi)}&\text{if}~~  y_i=1.\\
\end{cases}
\end{equation}

Concerning the M-step, we note that the log-likelihood in \eqref{ZIGEMll}
splits into two terms which can be maximized separately. The first term
contains the parameters of the extra zero model part ($\boldsymbol{\gamma}$) and the
other contains the parameters of the censored gamma distribution ($\alpha$
and $\boldsymbol{\beta}$).

\section{An Application}\label{SecAppl}
\subsection{Loss Given Default Data}
We apply the models presented above to a dataset from insurance. A surety bond is a contractual agreement among three parties: the
contractor who performs an obligation, the obligee who receives
the obligation, and the surety provider, in our case the insurance company, who
covers the risk that the contractor fails to fulfill the obligation.

The dataset consists of European surety bonds that resulted in a
claim. The ultimate loss for these claims is called ``Loss Given Default''
(LGD). For each bond, the maximal amount that is covered by the
insurance company, a quantity called ``face value'' (FV), is a priori
determined. This allows us to standardize
the LGD by dividing it by the face value, such that our variable of interest lies
between 0 and 1
\begin{equation}
0\leq \frac{\text{LGD}}{\text{FV}} \leq 1.
\end{equation}

We have worked with the original dataset, but for confidentiality reasons the
results presented here are obtained on the basis of a subsample of the
original set. The subsample, consisting of more than 5000 bonds, is
obtained by using a random selection mechanism, with selection
probabilities that depend on certain characteristics of the respective bonds, so that the value of the
average standardized loss LGD/FV is altered in order not to reveal the true average. As a consequence, the results presented in this paper are not
the real ones but are close enough to reflect the major
phenomena. We assure that the fit the models provide to the original
data is at least as good as for the subsample.

\begin{figure}[h]
\centering
\includegraphics[scale=0.4]{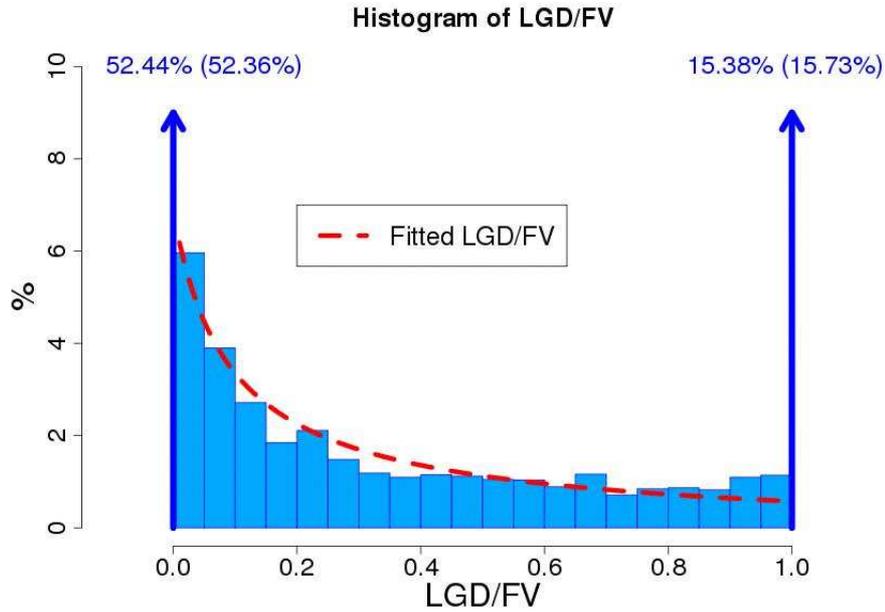} 
\caption{Histogram of LGD/FV and fitted censored gamma model with no
  covariates. The numbers above the blue arrows represent the
  percentage of LGD/FV's being exactly zero or one, respectively. In
  parentheses are the corresponding numbers as predicted by the censored gamma
  model. The dashed red line represents the fitted model.} 
\label{fig:CGnoCov}
\end{figure}

The standardized losses are shown in Figure \ref{fig:CGnoCov}. Since the insurance company can often recover costs,
observations with no ultimate loss at all are frequent. In fact, about $52
\%$ of all bonds in the subsample have no loss. On the other hand, there is a major
proportion ($15 \%$) of bonds that have full loss, i.e., a LGD/FV equaling 1.

Apart from providing a probabilistic model for the surety LGD, the purpose
is also to explore the relation of the losses
to certain covariates which are shortly described in the following.

The relative default time
(RDT) of a bond is the proportion of time that has passed at default since
its issuance over
the total life span of a bond. This quantity allows us to explore the time
development of the losses from the issuing date to the end date
(maturity). Experience and size are two categorical
variables, each attaining three different levels, which represent the
experience (low, mid, high) and the size (small, medium, large) of the
contractor. There
are three different types of surety bonds called
maintenance, performance, and hybrid bonds. Hybrid bonds are bonds that
are both maintenance and performance bonds. There is an additional category
denoted ``other bonds'' for a small number of bonds of various other categories. Usually,
European surety bonds do not cover the whole amount of an underlying
contract but only a certain fraction. Information about his
percentage is included as an additional covariate.   In Table
\ref{tab:SumStat} in the online supplementary material, we
report summary statistics for the continuous covariates and relative
frequencies for the categorical variables.  

\subsection{Results}\label{ressec}
\begin{figure}[h]
\centering
\includegraphics[scale=0.4,angle=0]{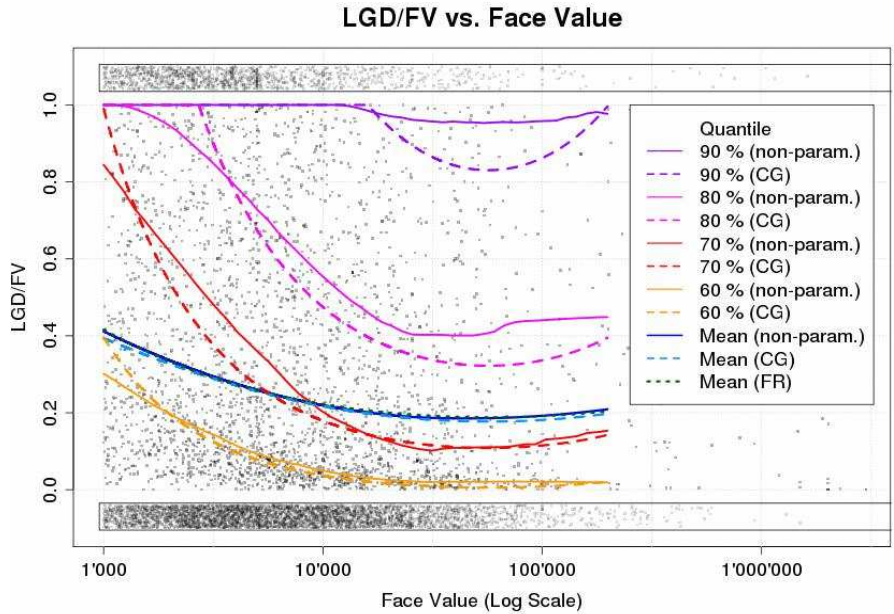} 
\caption{Scatter plot of face value (on a logarithmic scale)
  vs. LGD/FV. The jittered points in the bars below $0.0$ and above $1.0$ represent bonds with
  LGD/FV being exactly zero and one, respectively. The colored solid lines
  are non-parametrically fitted quantiles and mean. The dashed lines
  represent quantiles and mean of the fitted censored
  gamma (CG) model. The green dotted line represents the fitted conditional mean of the fractional response (FR) model. Logarithmic and squared logarithmic
  face value are taken as covariates.\label{LGDvsFVfit}}
\label{fig:CGFVCov}
\end{figure}

We first estimate the censored gamma model of Section \ref{SecCGMod} with no covariates and
illustrate its fit in Figure \ref{fig:CGnoCov}. The dashed red line
represents the fitted model. The numbers in parentheses above the bars show the
fitted probabilities of being zero and one. Apparently, the plain model with no
covariates fits the data well. The observed and the modeled
probabilities of being zero or one are very similar and the
continuous part of the model accurately fits the histogram.\footnote{Due to
  the large number of observations, a chi-square goodness of fit test still
  shows significant deviations.}   For comparison, we have also fitted the standard
normal Tobit model in its two-sided version, as well as a corresponding
model using a skewed t distribution (\citet{AzCa03}). See the  
supplementary material for more details.  Both models provide worse fits than the censored gamma model. A plot
(Figure \ref{fig:TobitnoCov})
illustrating the fits can be found in the   supplementary material.

Next, we fit a model using only the face value, more specifically the
logarithm and the squared logarithm of the face value, as covariate. We
illustrate the fitted model in Figure \ref{fig:CGFVCov}. The colored
continuous lines are non-parametrically fitted quantile (see
\citet{KoeR05}) and mean curves (calculated using local polynomial regression, see
\citet{ChaJH92}, Chapter 8). The dashed lines
  represent the corresponding quantiles and mean of the fitted model
  calculated using the result in Lemma \ref{LemIntp}. We also fit the conditional mean model
  for fractional response (FR) of \citet{PaWo96}. Here, fitting is done
  using quasi-maximum likelihood (see \citet{GoMoTr84} for details) based on the Bernoulli
  log-likelihood function. 

The non-parametrically fitted mean and the
mean of the fitted censored gamma model are very close
  together. This indicates that the censored gamma model provides a good
  fit to the conditional mean. Moreover, the non-parametrically estimated quantiles and the
  quantiles from the fitted censored gamma model match well. I.e., the censored gamma
  model not only models the mean appropriately but the entire
  distribution.  In addition, the fitted mean of fractional response model
  is very close the mean of the fitted censored gamma model.   Again, we have
  also fitted the Tobit model and the skewed t version. Compared to the
  censored gamma model, both models provide worse fits (see Figure
  \ref{fig:TobitFVCov} in the   supplementary material).  

Finally, we fit the censored gamma and its two extensions, i.e., the two-tiered
and the zero-inflated model including all
covariates. For the two ordinal factorial
variables experience and size, we use orthogonal polynomial
contrasts. Concerning the categorical variable type, we use treatment
contrasts with maintenance as baseline level. For the censored gamma model, we use
the Fisher scoring algorithm presented above. In the case of the two-tiered
and zero-inflated models, we use the algorithms presented in this paper to determine good
starting values for quasi-Newton methods. Starting values for the
parameters that are not estimated with these methods, i.e., the shape
parameter $\alpha$ and the location parameter $\xi$, respectively, are obtained by taking
the values from the ones in the fitted censored gamma model. We then
estimate the two models using quasi-Newton methods. Concerning the censored
gamma model, estimates of standard errors are
calculated using the Fisher information. For the other two models, standard errors are obtained by numerically
approximating the Fisher Information Matrix at the optimum. 

The results are reported in Table \ref{tab:fitfull}. The log-likelihood of both the two-tiered and zero-inflated models are
considerably higher than the one of the censored gamma model. This is also
reflected in considerably smaller AIC values, the
zero-inflated model having the lowest AIC.   A likelihood ratio test
clearly favors the two-tiered model over the censored gamma model. This is
also true for the zero-inflated model. For the latter, the null hypothesis
is on the boundary of the parameter space, and the usual asymptotics do not
apply. We therefore use a simulated test instead. To be more specific, the distribution of the
difference in log-likelihoods between the two models under the null
hypothesis is characterized by 1000 simulated values. A sample from this distribution is generated by
simulating data from the null hypothesis, i.e., from the estimated censored
Gamma model, then fitting both models, and calculating
the difference in the two log-likelihoods. The lowest simulated
difference obtained out of the 1000 samples was about $28.6$. We conclude that
the observed difference of more than 200 is clearly significant. Next, for discriminating between the two extended
models, we apply Vuong's test (\citet{Vu89}). Since we know that the
zero-inflated model does not reduce to the censored gamma model, it follows
that we are not in the overlapping case. Thus, we can use
the Vuong's non-nested hypothesis test.   The test statistic has a value of $-2.26$ under the
null hypothesis that both models are equally close to the true model. Thus,
at a $5\%$ level,
the null hypothesis is rejected in favor of the zero-inflated
model. This gives support to the idea that there are indeed extra zeros
in the data. These extra zeros are interpreted as zero losses from claims
that were filed for administrative reasons and not because there was a true
default event.   As before, we have also fitted the Tobit model and skewed t
distribution model using all covariates. The Results are reported in Tables
\ref{tab:fitTobitfull} and \ref{tab:fitSkewTfull} in the   supplementary material. In all cases, the gamma
models have considerably lower AICs, and the corresponding differences in
log-likelihood are always larger than 100, except when comparing the
two-tiered gamma model with the two-tiered skewed t model where the
differences is about 8 in favor of the gamma model. This means that Vuong's test favors the gamma model in all cases.


\begin{sidewaystable}[ht]
\begin{center}
\begin{tabular}{ll|rl|rlrl|rlrl}
\hline\hline
&  \multicolumn{1}{r|}{Model} &\multicolumn{2}{c}{Censored}&\multicolumn{4}{|c|}{Two-Tiered}&\multicolumn{4}{c}{Zero-Inflated}\\ 
Covariate & &  Coef & Std. Err. & Coef($\boldsymbol{\beta}$) & Std. Err. &Coef($\boldsymbol{\gamma}$) & Std. Err. & Coef($\boldsymbol{\beta}$) & Std. Err. &Coef($\boldsymbol{\gamma}$) & Std. Err. \\
\hline
 Intercept & & 3.9 & 0.34 *** & 3.9 & 0.33 *** & -3.2 & 0.61 *** & 4.1 & 0.35 *** & 0.023 & 0.18   \\
 \hline \multirow{2}{*}{RDT} &\text{Lin} & -0.17 & 0.10 $\cdot$ & 0.30 & 0.10 ** & -0.45 & 0.079 *** & 0.29 & 0.10 ** & 0.35 & 0.057 *** \\
   &\text{Quad} & 0.074 & 0.35   & 1.6 & 0.35 *** & -1.1 & 0.26 *** & 1.6 & 0.35 *** & 0.88 & 0.20 *** \\
 \hline \multirow{2}{*}{Experience} &\text{Lin} & -0.82 & 0.076 *** & -0.39 & 0.064 *** & -0.67 & 0.066 *** & -0.38 & 0.065 *** & 0.42 & 0.037 *** \\
   &\text{Quad} & 0.12 & 0.051 * & 0.064 & 0.045   & 0.068 & 0.041 $\cdot$ & 0.059 & 0.046   & -0.017 & 0.026   \\
 \hline \multirow{2}{*}{Size} &\text{Lin} & 0.56 & 0.32 $\cdot$ & 0.35 & 0.37   & 0.44 & 0.24 $\cdot$ & 0.34 & 0.38   & -0.36 & 0.19 $\cdot$ \\
   &\text{Quad} & 0.66 & 0.20 ** & -0.17 & 0.24   & 0.85 & 0.15 *** & -0.18 & 0.24   & -0.68 & 0.12 *** \\
 \hline \multirow{2}{*}{Face Value} &\text{Lin} & -0.80 & 0.071 *** & -0.96 & 0.065 *** & -0.0048 & 0.050   & -0.99 & 0.070 *** & -0.054 & 0.047   \\
   &\text{Quad} & 0.50 & 0.068 *** & 0.15 & 0.053 ** & 0.49 & 0.064 *** & 0.18 & 0.054 ** & -0.33 & 0.043 *** \\
 \hline \multirow{3}{*}{Type} & Hybrid & 2.9 & 1.5 $\cdot$ & 2.0 & 1.2 $\cdot$ & 2.7 & 1.2 * & 1.7 & 1.1   & -1.9 & 0.80 * \\
   & Performance & 0.015 & 0.12   & 0.16 & 0.11   & -0.12 & 0.099   & 0.17 & 0.11   & 0.12 & 0.070 $\cdot$ \\
   & Other & 0.23 & 0.16   & 0.52 & 0.17 ** & -0.20 & 0.12   & 0.57 & 0.17 ** & 0.19 & 0.095 * \\
 \hline Ins. Frac. & & 1.2 & 0.56 * & 1.5 & 0.49 ** & -0.43 & 0.39   & 1.6 & 0.49 *** & 0.40 & 0.28   \\
  \hline &  & Value & Std. Err.   &  & Value & Std. Err. &  &  &Value &  Std. Err.& \\
 \hline \multirow{2}{*}{Gamma Par.} &$\log(\alpha)$ & -1.5 & 0.050   & & -0.54 & 0.067   & & & -0.57 & 0.073   & \\
   & $\log(\xi)$ & -2.4 & 0.093   & & -4.5 & 0.47   & & & -4.3 & 0.44   & \\
\hline
\multicolumn{2}{c|}{Log-Likelihood}  &  \multicolumn{2}{c}{ -7898.4 } &\multicolumn{4}{|c|}{ -7684.9 } &\multicolumn{4}{c}{ -7680.5 } \\ 
\hline
\multicolumn{2}{c|}{AIC}  &  \multicolumn{2}{c}{ 15826.8 } &\multicolumn{4}{|c|}{ 15425.9 } &\multicolumn{4}{c}{ 15417.1 } \\ 
\hline
\end{tabular}
\end{center}
\caption{Fitted censored, two-tiered, and zero-inflated gamma models including all covariates. Codes for significance levels:   '***': $p<0.001$,  '**': $0.001\leq p < 0.01$,  '*':  $0.01 \leq p < 0.05$, '.': $0.05 \leq p < 0.1$.} \label{tab:fitfull}
\end{sidewaystable}

\subsection{Interpretation of Results}
 
Having come to the conclusion that the zero-inflated model provides the
best fit to our data, we interpret the obtained results. Interpretation is
not as straightforward as, for instance, in
the basic censored gamma model case (see Section \ref{CGInterpr}). In contrast to
that, in the zero-inflated extension there are two linear predictors
$\eta=\boldsymbol{x}'\boldsymbol{\beta}$ and
$\mu=\boldsymbol{x}'\boldsymbol{\gamma}$. Partial effects on, say, the
conditional mean therefore include both sets of coefficients $\boldsymbol{\beta}$
and $\boldsymbol{\gamma}$. We will focus on $E[Y|\boldsymbol{x}]$ and
$P[Y=0|\boldsymbol{x}]$ in the following. These two quantities and their corresponding
partial effects are calculated in the following lemma.

\begin{lem}\label{LemIntpZI}
For the zero-inflated model, the following relations hold true. 

\begin{equation}
E[Y|\boldsymbol{x}]=(1-\Phi(\mu))C^1_{\alpha,\vartheta,\xi}
\end{equation}
 where 
\begin{equation}
\begin{split}
C^1_{\alpha,\vartheta,\xi}=&\alpha\vartheta\left(G_{\alpha+1,\vartheta}(1+\xi)-G_{\alpha+1,\vartheta}(\xi)\right),\\
&+(1+\xi)\left(1-G_{\alpha,\vartheta}(1+\xi)\right)-\xi\left(1-G_{\alpha,\vartheta}(\xi)\right),
\end{split}
\end{equation}
and 
\begin{equation}
P[Y=0|\boldsymbol{x}]=\Phi(\mu)+(1-\Phi(\mu))\cdot G_{\alpha,\vartheta}(\xi).
\end{equation}

For a continuous covariate $x_j$, we have
\begin{equation} 
\frac{\partial E[Y|\boldsymbol{x}]}{\partial x_j}=\beta_jC^2_{\alpha,\vartheta,\xi} (1-\Phi(\mu))-\gamma_j\phi(\mu)C^1_{\alpha,\vartheta,\xi},
\end{equation}
where
\begin{equation}
C^2_{\alpha,\vartheta,\xi}=\alpha\vartheta (G_{\alpha+1,\vartheta}(1+\xi)-G_{\alpha+1,\vartheta}(\xi)),
\end{equation}

and
\begin{equation} 
\frac{\partial P[Y=0|\boldsymbol{x}]}{\partial x_j}=-\beta_j\xi g_{\alpha,\vartheta}(\xi)(1-\Phi(\mu))+\gamma_j\phi(\mu)\left(1-G_{\alpha,\vartheta}(\xi)\right).
\end{equation}

\end{lem}

The lemma follows from \eqref{ZIModel} together with Lemma
\ref{LemIntp}. We see that the partial effects contain $\boldsymbol{\beta}$
and $\boldsymbol{\gamma}$, both entering in a non-linear manner and
interacting with each other. This follows from the fact that $\vartheta=\exp(\boldsymbol{x}'\boldsymbol{\beta})$ and $\mu=\boldsymbol{x}'\boldsymbol{\gamma}$. Because of this we came to the conclusion that interpretation
is best done in a graphical way. This is done as described in the
following.

\begin{figure}[h]
\centering
\includegraphics[scale=0.25,angle=0]{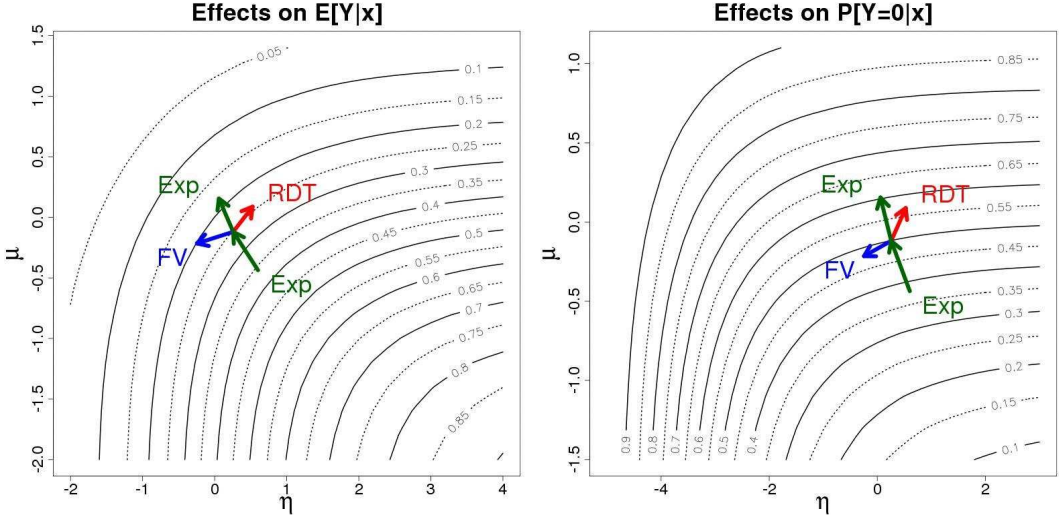} 
\caption{Illustration of effects of main covariates for the zero-inflated
  model. On the left hand side, a contour plot of the conditional
  expectation, $E[Y|\boldsymbol{x}]$, as a function of the two linear
  predictors $\eta=\boldsymbol{x}'\boldsymbol{\beta}$ and $\mu=\boldsymbol{x}'\boldsymbol{\gamma}$ is shown. On the right hand side, the same contour plot is shown
  for the probability of being zero, $P[Y=0|\boldsymbol{x}]$. The arrows
  represent the effects of changing covariates. For the two
  continuous covariates face value (FV) and relative default time (RDT),
  the arrows are obtained by increasing the variables by one standard
  deviation from their mean. For the factorial variable experience
  (Exp), the two arrows indicate the changes when moving from the lowest
  level to the middle one and then to the highest level.  \label{ZIEffects}}
\end{figure}

In Figure \ref{ZIEffects}, contour plots of the conditional
  expectation, $E[Y|\boldsymbol{x}]$, and the probability of being zero,
  $P[Y=0|\boldsymbol{x}]$, for the fitted zero-inflated model are shown. Contour levels are obtained
  with respect to varying values of the two linear predictors
  $\eta=\boldsymbol{x}'\boldsymbol{\beta}$ and
  $\mu=\boldsymbol{x}'\boldsymbol{\gamma}$. The arrows
  represent the effects of the covariates. The middle point of the arrows are the levels of
  $E[Y|\boldsymbol{x}]$ and $P[Y=0|\boldsymbol{x}]$, respectively,
  attained when taking all continuous covariates at their mean and the
  categorical variables at their most frequent level. We focus on the
  three variables face value (FV), relative default time (RDT), and experience
  (Exp) since these are believed to be the most important variables from a
  practical point of view. Interpretation for the other covariates is analogous. For the two
  continuous covariates face value (FV) and relative default time (RDT),
  the blue and red arrows in Figure \ref{ZIEffects} are obtained by increasing the variables by one standard
  deviation from their mean. For the categorical variable experience
  (Exp), the green arrows illustrate the changes in  $E[Y|\boldsymbol{x}]$ and $P[Y=0|\boldsymbol{x}]$ when moving from the lowest
  level to the middle one and then to the highest level of experience.

Concerning the conditional expectation, the blue arrow of the FV shows that an
increase of FV by one standard deviation leads to an increase in
$E[Y|\boldsymbol{x}]$ by about $0.05$. RDT, on the other
hand, has virtually no effect on the mean. Even though both linear
predictors change considerably when increasing RDT, the change is along a contour level and has no effect on the
value of $E[Y|\boldsymbol{x}]$. Concerning the experience, we observe
strong effects when going from low experience to middle and high, with a
total decrease of about $0.17$. 

For the probability of being zero, the picture is slightly different. FV
has only a small effect on $P[Y=0|\boldsymbol{x}]$, whereas increasing RDT by one standard
deviation results in an increase of about $6 \%$ in $P[Y=0|\boldsymbol{x}]$. Experience again has a
strong effect. $P[Y=0|\boldsymbol{x}]$ increases by more than $20 \%$ when
going from low to high experience.

 
\section{Conclusion}
Three special regression models for fractional response variables that attain their boundaries frequently were
presented. The first model determines the distribution of the values between the limits and the frequency of the
limiting values in a parsimonious way. Two extensions of this model to cover cases in which the
frequencies of the limits do not follow this parsimonious description were
introduced as well. The models were applied to a LGD dataset from
insurance. They were found to fit the data in a specific insurance
application better than other popular parametric models.


\section*{Acknowledgments}
We thank Hans-Rudolf K\"unsch for helpful comments and discussions. We
would also like to thank two anonymous referees for their insightful
comments and suggestions.

\clearpage

\begin{appendix}
\section{Proof of Lemma 2.1}\label{ProofLem}
Firstly, a censored gamma
distribution with density as in \eqref{CGdens2} has expectation
\begin{align}
E[Y|\boldsymbol{x}]=&0\cdot G_{\alpha,\vartheta}(\xi)+ \int_0^1{yg_{\alpha,\vartheta}(y+\xi)dy}+1\cdot
(1-G_{\alpha,\vartheta}(1+\xi))\nonumber \\
=&\int_{\xi}^{1+\xi}{(z-\xi)g_{\alpha,\vartheta}(z)dz}+(1-G_{\alpha,\vartheta}(1+\xi))\nonumber\\
=&\alpha\vartheta(G_{\alpha+1,\vartheta}(1+\xi)-G_{\alpha+1,\vartheta}(\xi))+\xi G_{\alpha,\vartheta}(\xi)\nonumber\\
&-\xi G_{\alpha,\vartheta}(1+\xi)+(1-G_{\alpha,\vartheta}(1+\xi))\nonumber\\
=&\alpha\vartheta(G_{\alpha+1,\vartheta}(1+\xi)-G_{\alpha+1,\vartheta}(\xi))\nonumber\\
&+(1+\xi)\left(1-G_{\alpha,\vartheta}(1+\xi)\right)-\xi\left(1-G_{\alpha,\vartheta}(\xi)\right), \label{CGmean}
\end{align}
where in the third line we have used the identity \eqref{expcengam} given in the supplementary material.

Secondly, for a continuous $x_j$, using 
\begin{equation}\label{gammaDerivs}
\frac{\partial G_{\alpha,\vartheta}(\xi)}{\partial \vartheta} =
\frac{\partial G_{\alpha,1}(\xi/\vartheta)}{\partial \vartheta}=-\frac{\xi}{\vartheta^2}g_{\alpha,1}\left(\frac{\xi}{\vartheta}\right)=-\frac{\xi}{\vartheta}g_{\alpha,\vartheta}\left(\xi\right),
\end{equation}
or
\begin{equation}\label{gammaDerivs2}
\frac{\partial G_{\alpha,\vartheta}(\xi)}{\partial \vartheta}=-\alpha g_{\alpha+1,\vartheta}\left(\xi\right),
\end{equation}
 and the fact that $$\frac{\partial \vartheta}{\partial x_j}=\vartheta \beta_j,$$ we can compute the partial derivatives of $E[Y|\boldsymbol{x}]$ with
respect to $x_j$ as
\begin{align}
\frac{\partial E[Y|\boldsymbol{x}]}{\partial
  x_j}=&-\xi \alpha g_{\alpha+1,\vartheta}\left(\xi\right)\vartheta \beta_j+\alpha\vartheta \beta_j(G_{\alpha+1,\vartheta}(1+\xi)-G_{\alpha+1,\vartheta}(\xi))\nonumber\\
&-\alpha\vartheta\frac{1+\xi}{\vartheta}g_{\alpha+1,\vartheta}\left(1+\xi\right)\vartheta \beta_j+\alpha\vartheta\frac{\xi}{\vartheta}g_{\alpha+1,\vartheta}\left(\xi\right)\vartheta \beta_j\nonumber\\
&+(1+\xi)\alpha g_{\alpha+1,\vartheta}\left(1+\xi\right)\vartheta \beta_j\nonumber\\
=&\alpha\vartheta(G_{\alpha+1,\vartheta}(1+\xi)-G_{\alpha+1,\vartheta}(\xi))\beta_j.
\end{align}
\end{appendix}

\clearpage

\bibliographystyle{plainnat}
\bibliography{FSbiblio}

\vspace{2cm}

\noindent
Fabio Sigrist (corresponding author)\\
Seminar for Statistics, Department of Mathematics\\
ETH Z\"urich\\
R\"amistrasse 110\\
CH-8092 Z\"urich\\
Switzerland\\
E-Mail: sigrist@stat.math.ethz.ch\\

\noindent
Werner A. Stahel\\
Seminar for Statistics, Department of Mathematics\\
ETH Z\"urich\\
R\"amistrasse 110\\
CH-8092 Z\"urich\\
Switzerland\\
E-Mail: stahel@stat.math.ethz.ch\\

\clearpage 

\section*{Supplementary Material}
\renewcommand{\thesubsection}{S.\arabic{subsection}}
\subsection{Fisher Information Matrix for the Censored Gamma Model}\label{app1}
In the following derivations, we will often use some identities and results on integrals that we list in Section \ref{IdInt} below.

With \eqref{CGllDera}, it follows that
\begin{equation*}
  \begin{split}
E_{\theta}&\left[ \frac{\partial \ell_i}{\partial \alpha'}\frac{\partial
    \ell_i}{\partial \alpha'}\right]\\
&=E_{\theta}\left[\left(\frac{\alpha}{G_{\alpha,\vartheta_i}(\xi)}\left(-\psi(\alpha)G_{\alpha,\vartheta_i}(\xi) +  H^{(1)}_{\alpha}\left(0, \frac{\xi}{\vartheta_i}\right)\right)\mathbf{1}_{\{y_i=0\}}\right)^2\right]\\ &+E_{\theta}\left[\left(\alpha\left(-\log(\vartheta_i)-\psi(\alpha)+\log(y_i+\xi)\right)\mathbf{1}_{\{0<  y_i <  1\}}\right)^2\right]\\ &+E_{\theta}\left[\left(-\frac{\alpha}{1-G_{\alpha,\vartheta_i}(1+\xi)}\left(-\psi(\alpha)G_{\alpha,\vartheta_i}(1+\xi) +
  H^{(1)}_{\alpha}\left(0, \frac{1+\xi}{\vartheta_i}\right)\right)\mathbf{1}_{\{y_i=1\}} \right)^2\right]\\
&=\left(\frac{\alpha}{G_{\alpha,\vartheta_i}(\xi)}\left(-\psi(\alpha)G_{\alpha,\vartheta_i}(\xi) +
    H^{(1)}_{\alpha}\left(0, \frac{\xi}{\vartheta_i}\right)\right)\right)^2\cdot G_{\alpha,\vartheta_i}(\xi)\\ &+\int_0^1{\left(\alpha\left(-\log(\vartheta_i)-\psi(\alpha)+\log(y_i+\xi)\right)\right)^2g_{\alpha,\vartheta_i}(y_i+\xi)dy_i}\\ &+\left(\frac{\alpha}{1-G_{\alpha,\vartheta_i}(1+\xi)}\left(-\psi(\alpha)G_{\alpha,\vartheta_i}(1+\xi) +
  H^{(1)}_{\alpha}\left(0, \frac{1+\xi}{\vartheta_i}\right)\right)\right)^2\cdot (1-G_{\alpha,\vartheta_i}(1+\xi)).\\
\end{split}
\end{equation*}
Using \eqref{expcengamlog} and \eqref{expcengamlog2}, the middle summand of
this expression is calculated as
\begin{equation*}
  \begin{split}
\int_0^1&{\left(\alpha\left(-\log(\vartheta_i)-\psi(\alpha)+\log(y_i+\xi)\right)\right)^2g_{\alpha,\vartheta_i}(y_i+\xi)dy_i}
\\ =&\alpha^2(\log(\vartheta_i)+\psi(\alpha))^2(G_{\alpha,\vartheta_i}(1+\xi)-G_{\alpha,\vartheta_i}(\xi))\\
&-2\alpha^2(\log(\vartheta_i)+\psi(\alpha))\left(\log(\vartheta_i)(G_{\alpha,\vartheta_i}(1+\xi)-G_{\alpha,\vartheta_i}(\xi))+H^{(1)}_{\alpha}\left(\frac{\xi}{\vartheta_i},\frac{1+\xi}{\vartheta_i}\right)\right)\\
&+\alpha^2\log(\vartheta_i)^2(G_{\alpha,\vartheta_i}(1+\xi)-G_{\alpha,\vartheta_i}(\xi))+2\alpha^2\log(\vartheta_i)H^{(1)}_{\alpha}\left(\frac{\xi}{\vartheta_i},\frac{1+\xi}{\vartheta_i}\right)\\
&+\alpha^2H^{(2)}_{\alpha}\left(\frac{\xi}{\vartheta_i},\frac{1+\xi}{\vartheta_i}\right)\\
=&\alpha^2\psi(\alpha)^2(G_{\alpha,\vartheta_i}(1+\xi)-G_{\alpha,\vartheta_i}(\xi))-2\alpha^2\psi(\alpha)H^{(1)}_{\alpha}\left(\frac{\xi}{\vartheta_i},\frac{1+\xi}{\vartheta_i}\right)+\alpha^2H^{(2)}_{\alpha}\left(\frac{\xi}{\vartheta_i},\frac{1+\xi}{\vartheta_i}\right).
\end{split}
\end{equation*}
From this follows that
\begin{equation*}
  \begin{split}
E_{\theta}&\left[ \frac{\partial \ell_i}{\partial \alpha'}\frac{\partial \ell_i}{\partial \alpha'}
 \right]\\
=&\frac{\alpha^2}{G_{\alpha,\vartheta_i}(\xi)}\left(-\psi(\alpha)G_{\alpha,\vartheta_i}(\xi) +  H^{(1)}_{\alpha}\left(0,
    \frac{\xi}{\vartheta_i}\right)\right)^2\\
&+\alpha^2\left(\psi(\alpha)^2(G_{\alpha,\vartheta_i}(1+\xi)-G_{\alpha,\vartheta_i}(\xi))-2\psi(\alpha)H^{(1)}_{\alpha}\left(\frac{\xi}{\vartheta_i},\frac{1+\xi}{\vartheta_i}\right)+H^{(2)}_{\alpha}\left(\frac{\xi}{\vartheta_i},\frac{1+\xi}{\vartheta_i}\right)\right)\\
&+\frac{\alpha^2}{1-G_{\alpha,\vartheta_i}(1+\xi)}\left(-\psi(\alpha)G_{\alpha,\vartheta_i}(1+\xi) +
  H^{(1)}_{\alpha}\left(0, \frac{1+\xi}{\vartheta_i}\right)\right)^2.
\end{split}
\end{equation*}

For the remaining entries of the Fisher Information Matrix, the
calculation procedure is similar to the one made before. That is, the
computation of each expectation can be split in to three terms of which the middle term,
corresponding to the non-censored part of the model, requires more effort
to compute. In the following, we therefore first calculate the corresponding
middle term in each case.
 
With \eqref{expcengam}, \eqref{expcengamlog},
\eqref{expcengamgamlog}, and \eqref{gammaidentity2}, we calculate
\begin{equation*}
  \begin{split}
E_{\theta}&\left[\alpha\left(-\log(\vartheta_i)-\psi(\alpha)+\log(y_i+\xi)\right)x_{ik}\left(-\alpha+\frac{y_i+\xi}{\vartheta_i}\right)  \mathbf{1}_{\{0<  y_i <  1\}}\right] \\ 
=&\alpha^2x_{ik}\log(\vartheta_i)(G_{\alpha,\vartheta_i}(1+\xi)-G_{\alpha,\vartheta_i}(\xi))
+\alpha^2x_{ik}\psi(\alpha)(G_{\alpha,\vartheta_i}(1+\xi)-G_{\alpha,\vartheta_i}(\xi))\\
&-\alpha^2x_{ik}\log(\vartheta_i)(G_{\alpha+1,\vartheta_i}(1+\xi)-G_{\alpha+1,\vartheta_i}(\xi))-\alpha^2x_{ik}\psi(\alpha)(G_{\alpha+1,\vartheta_i}(1+\xi)-G_{\alpha+1,\vartheta_i}(\xi))\\
&-\alpha^2x_{ik}\log(\vartheta_i)(G_{\alpha,\vartheta_i}(1+\xi)-G_{\alpha,\vartheta_i}(\xi))-\alpha^2x_{ik}H^{(1)}_{\alpha}\left(\frac{\xi}{\vartheta_i},\frac{1+\xi}{\vartheta_i}\right)\\
&+\alpha^2x_{ik}\log(\vartheta_i)(G_{\alpha+1,\vartheta_i}(1+\xi)-G_{\alpha+1,\vartheta_i}(\xi))+\alpha^2x_{ik}H_{\alpha+1}\left(\frac{\xi}{\vartheta_i},\frac{1+\xi}{\vartheta_i}\right)\\
=&\alpha^2x_{ik}\psi(\alpha)(G_{\alpha+1,\vartheta_i}(\xi)-G_{\alpha,\vartheta_i}(\xi)-G_{\alpha+1,\vartheta_i}(1+\xi)+G_{\alpha,\vartheta_i}(1+\xi))\\
&+\alpha^2x_{ik}\left(-H^{(1)}_{\alpha}\left(\frac{\xi}{\vartheta_i},\frac{1+\xi}{\vartheta_i}\right)+H_{\alpha+1}\left(\frac{\xi}{\vartheta_i},\frac{1+\xi}{\vartheta_i}\right)\right)\\
=&\alpha^2x_{ik}\left(\psi(\alpha)
    \vartheta_ig_{\alpha+1,\vartheta_i}\left(1+\xi\right)-\psi(\alpha)\vartheta_ig_{\alpha+1,\vartheta_i}\left(\xi\right)\right)\\
  &-\alpha^2x_{ik}\left(H^{(1)}_{\alpha}\left(\frac{\xi}{\vartheta_i},\frac{1+\xi}{\vartheta_i}\right)+H_{\alpha+1}\left(\frac{\xi}{\vartheta_i},\frac{1+\xi}{\vartheta_i}\right)\right).
\end{split}
\end{equation*}
Using this result, \eqref{CGllDera}, and \eqref{CGllDerb}, we get
\begin{equation*}
  \begin{split}
  E_{\theta}&\left[ \frac{\partial \ell_i}{\partial
      \alpha'}\frac{\partial \ell_i}{\partial \beta_k}  \right]\\  = 
& E_{\theta}\left[\frac{\alpha}{G_{\alpha,\vartheta_i}(\xi)}\left(-\psi(\alpha)G_{\alpha,\vartheta_i}(\xi) +  H^{(1)}_{\alpha}\left(0, \frac{\xi}{\vartheta_i}\right)\right)\frac{-x_{ik}\xi\cdot g_{\alpha,\vartheta_i}\left(\xi\right)}{
  G_{\alpha,\vartheta_i}(\xi)}\mathbf{1}_{\{y_i=0\}} \right]\\ 
&+E_{\theta}\left[\alpha\left(-\log(\vartheta_i)-\psi(\alpha)+\log(y_i+\xi)\right)x_{ik}\left(-\alpha+\frac{y_i+\xi}{\vartheta_i}\right)\mathbf{1}_{\{0<  y_i <  1\}} \right]\\ 
&+E_{\theta}\left[\frac{-\alpha\left(-\psi(\alpha)G_{\alpha,\vartheta_i}(1+\xi) +
  H^{(1)}_{\alpha}\left(0,
    \frac{1+\xi}{\vartheta_i}\right)\right)}{1-G_{\alpha,\vartheta_i}(1+\xi)}\frac{x_{ik}(1+\xi)\cdot
g_{\alpha,\vartheta_i}\left(1+\xi\right)}{1-
G_{\alpha,\vartheta_i}(1+\xi)}\mathbf{1}_{\{y_i=1\}} \right]\\
=&-x_{ik}\frac{a\xi\cdot
  g_{\alpha,\vartheta_i}\left(\xi\right)\left(-\psi(\alpha)G_{\alpha,\vartheta_i}(\xi) +
    H^{(1)}_{\alpha}\left(0, \frac{\xi}{\vartheta_i}\right)\right)}{G_{\alpha,\vartheta_i}(\xi)}\\
&+x_{ik}\alpha^2\left(\psi(\alpha)
    \vartheta_ig_{\alpha+1,\vartheta_i}\left(\xi+1\right)-\psi(\alpha)\vartheta_ig_{\alpha+1,\vartheta_i}\left(\xi\right)\right)\\
  &-x_{ik}\alpha^2\left(H^{(1)}_{\alpha}\left(\frac{\xi}{\vartheta_i},\frac{1+\xi}{\vartheta_i}\right)+H_{\alpha+1}\left(\frac{\xi}{\vartheta_i},\frac{1+\xi}{\vartheta_i}\right)\right) \\ 
&-x_{ik}\frac{a(1+\xi)\cdot
g_{\alpha,\vartheta_i}\left(1+\xi\right)\left(-\psi(\alpha)G_{\alpha,\vartheta_i}(1+\xi) +
  H^{(1)}_{\alpha}\left(0,
    \frac{1+\xi}{\vartheta_i}\right)\right)}{1-G_{\alpha,\vartheta_i}(1+\xi)}.
\end{split}
\end{equation*}
Next, with \eqref{expcengam}, \eqref{expcengam2}, and \eqref{gammaidentity2}, we calculate
\begin{equation*}
\begin{split}
E_{\theta}&\left[
  x_{ik}x_{il}\left(-\alpha+\frac{y_i+\xi}{\vartheta_i}\right)^2\mathbf{1}_{\{0<    y_i <  1\}}\right]\\
&=x_{ik}x_{il}\alpha^2(G_{\alpha,\vartheta_i}(1+\xi)-G_{\alpha,\vartheta_i}(\xi))-2\alpha^2x_{ik}x_{il}(G_{\alpha+1,\vartheta_i}(1+\xi)-G_{\alpha+1,\vartheta_i}(\xi))\\
&+a(\alpha+1)x_{ik}x_{il}(G_{\alpha+2,\vartheta_i}(1+\xi)-G_{\alpha+2,\vartheta_i}(\xi))\\
&=\alpha^2x_{ik}x_{il}\vartheta_i\left(g_{\alpha+1,\vartheta_i}\left(1+\xi\right)-g_{\alpha+1,\vartheta_i}\left(\xi\right)-g_{\alpha+2,\vartheta_i}\left(1+\xi\right)+g_{\alpha+2,\vartheta_i}\left(\xi\right)\right)\\
&+\alpha x_{ik}x_{il}(G_{\alpha+2,\vartheta_i}(1+\xi)-G_{\alpha+2,\vartheta_i}(\xi)).
\end{split}
\end{equation*}
Using this result and \eqref{CGllDerb}, we see that
\begin{equation*}
\begin{split}
  E_{\theta}&\left[ \frac{\partial \ell_i}{\partial \beta_k}\frac{\partial \ell_i}{\partial \beta_l}  \right] \\ = & E_{\theta}\left[x_{ik}x_{il} \left(\frac{\xi\cdot g_{\alpha,\vartheta_i}\left(\xi\right)}{  G_{\alpha,\vartheta_i}(\xi)}\right)^2 \mathbf{1}_{\{y_i=0\}} \right]\\
&+E_{\theta}\left[x_{ik}x_{il}\left(-\alpha+\frac{y_i+\xi}{\vartheta_i}\right)^2\mathbf{1}_{\{0<  y_i  < 1\}} \right]\\
&+E_{\theta}\left[x_{ik}x_{il}\left(\frac{(1+\xi)\cdot g_{\alpha,\vartheta_i}\left(1+\xi\right)}{1- G_{\alpha,\vartheta_i}(1+\xi)}\right)^2\mathbf{1}_{\{y_i=1\}}\right]\\
=&\alpha^2x_{ik}x_{il}\vartheta_i\left(g_{\alpha+1,\vartheta_i}\left(1+\xi\right)-g_{\alpha+1,\vartheta_i}\left(\xi\right)-g_{\alpha+2,\vartheta_i}\left(1+\xi\right)+g_{\alpha+2,\vartheta_i}\left(\xi\right)\right) \\
&+x_{ik}x_{il}\left(\alpha(G_{\alpha+2,\vartheta_i}(1+\xi)-G_{\alpha+2,\vartheta_i}(\xi))+\frac{\left(\xi\cdot g_{\alpha,\vartheta_i}\left(\xi\right)\right)^2}{ G_{\alpha,\vartheta_i}(\xi)}+\frac{\left((1+\xi)\cdot g_{\alpha,\vartheta_i}\left(1+\xi\right)\right)^2}{1-  G_{\alpha,\vartheta_i}(1+\xi)}\right).
\end{split}
\end{equation*}
Moreover, with \eqref{expcengam-1}, \eqref{expcengamlog},
\eqref{expcengam-1gamlog}, and \eqref{gammaidentity2}, we get
\begin{equation*}
\begin{split}
E_{\theta}&\left[\alpha\left(-\log(\vartheta_i)-\psi(\alpha)+\log(y_i+\xi)\right)\left(\frac{\alpha-1}{y_i+\xi}-\frac{1}{\vartheta_i}\right)  \mathbf{1}_{\{0<  y_i <  1\}} \right] \\ 
=&\frac{-\alpha\log(\vartheta_i)}{\vartheta_i}(G_{\vartheta_i,a-1}(1+\xi)-G_{\vartheta_i,a-1}(\xi))-\frac{\alpha\psi(\alpha)}{\vartheta_i}(G_{\vartheta_i,a-1}(1+\xi)-G_{\vartheta_i,a-1}(\xi))\\
& +\frac{\alpha\log(\vartheta_i)}{\vartheta_i}(G_{\alpha,\vartheta_i}(1+\xi)-G_{\alpha,\vartheta_i}(\xi))+\frac{\alpha\psi(\alpha)}{\vartheta_i}(G_{\alpha,\vartheta_i}(1+\xi)-G_{\alpha,\vartheta_i}(\xi))\\
&+\frac{\alpha\log(\vartheta_i)}{\vartheta_i}(G_{\vartheta_i,a-1}(1+\xi)-G_{\vartheta_i,a-1}(\xi))+\frac{\alpha}{\vartheta_i}H_{\alpha-1}\left(\frac{\xi}{\vartheta_i},\frac{1+\xi}{\vartheta_i}\right)\\
&-\frac{\alpha\log(\vartheta_i)}{\vartheta_i}(G_{\alpha,\vartheta_i}(1+\xi)-G_{\alpha,\vartheta_i}(\xi))-\frac{\alpha}{\vartheta_i}H_{\alpha}\left(\frac{\xi}{\vartheta_i},\frac{1+\xi}{\vartheta_i}\right)\\
=&\frac{\alpha\psi(\alpha)}{\vartheta_i}(G_{\alpha,\vartheta_i}(1+\xi)-G_{\vartheta_i,a-1}(1+\xi)-G_{\alpha,\vartheta_i}(\xi)+G_{\vartheta_i,a-1}(\xi))\\
&+\frac{\alpha}{\vartheta_i}\left(H_{\alpha-1}\left(\frac{\xi}{\vartheta_i},\frac{1+\xi}{\vartheta_i}\right)-H_{\alpha}\left(\frac{\xi}{\vartheta_i},\frac{1+\xi}{\vartheta_i}\right)\right)\\
=&\alpha\psi(\alpha)\left(-g_{\alpha,\vartheta_i}\left(\xi+1\right)+ g_{\alpha,\vartheta_i}\left(\xi\right)\right)+\frac{\alpha}{\vartheta_i}\left(H_{\alpha-1}\left(\frac{\xi}{\vartheta_i},\frac{1+\xi}{\vartheta_i}\right)-H_{\alpha}\left(\frac{\xi}{\vartheta_i},\frac{1+\xi}{\vartheta_i}\right)\right).
\end{split}
\end{equation*}
With this equation, \eqref{CGllDera}, and \eqref{CGllDerh}, we calculate
\begin{equation*}
  \begin{split}
  E_{\theta}&\left[ \frac{\partial \ell_i}{\partial
      \alpha'}\frac{\partial \ell_i}{\partial \xi}  \right]\\  = 
& E_{\theta}\left[\frac{\alpha}{G_{\alpha,\vartheta_i}(\xi)}\left(-\psi(\alpha)G_{\alpha,\vartheta_i}(\xi) +  H^{(1)}_{\alpha}\left(0, \frac{\xi}{\vartheta_i}\right)\right)\frac{g_{\alpha,\vartheta_i}(\xi)}{G_{\alpha,\vartheta_i}(\xi)}\mathbf{1}_{\{y_i=0\}} \right]\\ 
&+E_{\theta}\left[\alpha\left(-\log(\vartheta_i)-\psi(\alpha)+\log(y_i+\xi)\right)\left(\frac{\alpha-1}{y_i+\xi}-\frac{1}{\vartheta_i}\right)  \mathbf{1}_{\{0<  y_i <  1\}} \right] \\ 
&+E_{\theta}\left[\frac{-\alpha\left(-\psi(\alpha)G_{\alpha,\vartheta_i}(1+\xi) +
  H^{(1)}_{\alpha}\left(0,    \frac{1+\xi}{\vartheta_i}\right)\right)}{1-G_{\alpha,\vartheta_i}(1+\xi)}\frac{-g_{\alpha,\vartheta_i}(1+\xi)}{1-G_{\alpha,\vartheta_i}(1+\xi)}  \mathbf{1}_{\{y_i=1\}} \right]\\
=&\frac{\alpha\left(-\psi(\alpha)G_{\alpha,\vartheta_i}(\xi) +
    H^{(1)}_{\alpha}\left(0, \frac{\xi}{\vartheta_i}\right)\right)g_{\alpha,\vartheta_i}(\xi)}{G_{\alpha,\vartheta_i}(\xi)}\\
&+ \alpha\psi(\alpha)\left(-g_{\alpha,\vartheta_i}\left(\xi+1\right)+ g_{\alpha,\vartheta_i}\left(\xi\right)\right)+\frac{\alpha}{\vartheta_i}\left(H_{\alpha-1}\left(\frac{\xi}{\vartheta_i},\frac{1+\xi}{\vartheta_i}\right)-H_{\alpha}\left(\frac{\xi}{\vartheta_i},\frac{1+\xi}{\vartheta_i}\right)\right) \\ 
&+\frac{\alpha\left(-\psi(\alpha)G_{\alpha,\vartheta_i}(1+\xi) +
  H^{(1)}_{\alpha}\left(0,
    \frac{1+\xi}{\vartheta_i}\right)\right)g_{\alpha,\vartheta_i}(1+\xi)}{1-G_{\alpha,\vartheta_i}(1+\xi)}.
\end{split}
\end{equation*}
With \eqref{expcengam}, \eqref{expcengam-1}, and \eqref{gammaidentity2}, we calculate
\begin{equation*}
\begin{split}
E_{\theta}&\left[
  x_{ik}\left(-\alpha+\frac{y_i+\xi}{\vartheta_i}\right)\left(\frac{\alpha-1}{y_i+\xi}-\frac{1}{\vartheta_i}\right)\mathbf{1}_{\{0<    y_i <  1\}}\right]\\
=&\frac{-\alpha x_{ik}(G_{\vartheta_i,a-1}(1+\xi)-G_{\vartheta_i,a-1}(\xi))}{\vartheta_i}+\frac{\alpha x_{ik}(G_{\alpha,\vartheta_i}(1+\xi)-G_{\alpha,\vartheta_i}(\xi))}{\vartheta_i}\\
&+\frac{(\alpha-1)x_{ik}(G_{\alpha,\vartheta_i}(1+\xi)-G_{\alpha,\vartheta_i}(\xi))}{\vartheta_i}-\frac{\alpha x_{ik}(G_{\alpha+1,\vartheta_i}(1+\xi)-G_{\alpha+1,\vartheta_i}(\xi))}{\vartheta_i}\\
=&x_{ik}\alpha\left(-g_{\alpha,\vartheta_i}\left(1+\xi\right)+g_{\alpha,\vartheta_i}\left(\xi\right)+g_{\alpha+1,\vartheta_i}\left(1+\xi\right)-g_{\alpha+1,\vartheta_i}\left(\xi\right)\right)\\
&-\frac{x_{ik}(G_{\alpha,\vartheta_i}(1+\xi)-G_{\alpha,\vartheta_i}(\xi))}{\vartheta_i}.
\end{split}
\end{equation*}
Using the above result, we have
\begin{equation*}
\begin{split}
  E_{\theta}&\left[ \frac{\partial \ell_i}{\partial \beta_k}\frac{\partial \ell_i}{\partial \xi}  \right] \\ = & E_{\theta}\left[-x_{ik} \frac{\xi g_{\alpha,1}\left(\frac{\xi}{\vartheta_i}\right)}{  G_{\alpha,\vartheta_i}(\xi)}\frac{g_{\alpha,\vartheta_i}(\xi)}{G_{\alpha,\vartheta_i}(\xi)} \mathbf{1}_{\{y_i=0\}} \right]\\
&+E_{\theta}\left[x_{ik}\left(-\alpha+\frac{y_i+\xi}{\vartheta_i}\right)\left(\frac{\alpha-1}{y_i+\xi}-\frac{1}{\vartheta_i}\right)\mathbf{1}_{\{0<  y_i  < 1\}} \right]\\
&+E_{\theta}\left[x_{ik}\frac{(1+\xi) g_{\alpha,\vartheta_i}\left(1+\xi\right)}{1- G_{\alpha,\vartheta_i}(1+\xi)}\frac{-g_{\alpha,\vartheta_i}(1+\xi)}{1-G_{\alpha,\vartheta_i}(1+\xi)}\mathbf{1}_{\{y_i=1\}}\right]\\
=&x_{ik}\alpha\left(-g_{\alpha,\vartheta_i}\left(1+\xi\right)+g_{\alpha,\vartheta_i}\left(\xi\right)+g_{\alpha+1,\vartheta_i}\left(1+\xi\right)-g_{\alpha+1,\vartheta_i}\left(\xi\right)\right)\\
&+x_{ik}\left(-\frac{G_{\alpha,\vartheta_i}(1+\xi)-G_{\alpha,\vartheta_i}(\xi)}{\vartheta_i}-\frac{\xi g_{\alpha,\vartheta_i}(\xi)^2}{  G_{\alpha,\vartheta_i}(\xi)}-\frac{(1+\xi)\cdot g_{\alpha,\vartheta_i}(1+\xi)^2}{1-  G_{\alpha,\vartheta_i}(1+\xi)}\right).
\end{split}
\end{equation*}
Next, with \eqref{expcengam-1}, \eqref{expcengam-2},
\eqref{gammaidentity2}, we calculate
\begin{equation*}
\begin{split}
E_{\theta}&\left[\left(\frac{\alpha-1}{y_i+\xi}-\frac{1}{\vartheta_i}\right)^2\mathbf{1}_{\{0<    y_i <  1\}}\right]\\
=&\frac{(\alpha-1)(G_{\vartheta_i,a-2}(1+\xi)-G_{\vartheta_i,a-2}(\xi))}{(\alpha-2)\vartheta_i^2}-2\frac{G_{\vartheta_i,a-1}(1+\xi)-G_{\vartheta_i,a-1}(\xi)}{\vartheta_i^2}\\
&+\frac{G_{\alpha,\vartheta_i}(1+\xi)-G_{\alpha,\vartheta_i}(\xi)}{\vartheta_i^2}\\
=&\left(\frac{\vartheta_i(\alpha-1)^2-(\xi+1)(\alpha-3)}{\vartheta_i(\alpha-2)(\xi+1)}\right)g_{\alpha,\vartheta_i}\left(\xi+1\right)-\left(\frac{\vartheta_i(\alpha-1)^2-\xi (\alpha-3)}{\vartheta_i(\alpha-2)\xi}\right)
  g_{\alpha,\vartheta_i}\left(\xi\right)\\ &+\frac{G_{\alpha,\vartheta_i}(1+\xi)-G_{\alpha,\vartheta_i}(\xi)}{(\alpha-2)\vartheta_i^2}.
\end{split}
\end{equation*}
Finally, using this result, we have
\begin{equation*}
\begin{split}
  E_{\theta}&\left[ \frac{\partial \ell_i}{\partial \xi}\frac{\partial \ell_i}{\partial \xi}  \right] \\ = & E_{\theta}\left[\left(\frac{g_{\alpha,\vartheta_i}(\xi)}{G_{\alpha,\vartheta_i}(\xi)}\right)^2 \mathbf{1}_{\{y_i=0\}} \right]+E_{\theta}\left[\left(\frac{\alpha-1}{y_i+\xi}-\frac{1}{\vartheta_i}\right)^2\mathbf{1}_{\{0<  y_i  < 1\}} \right]\\
&+E_{\theta}\left[\left(\frac{-g_{\alpha,\vartheta_i}(1+\xi)}{1-G_{\alpha,\vartheta_i}(1+\xi)}\right)^2\mathbf{1}_{\{y_i=1\}}\right]\\
= &  \frac{g_{\alpha,\vartheta_i}(\xi)^2}{ G_{\alpha,\vartheta_i}(\xi)} + \left(\frac{\vartheta_i(\alpha-1)^2-(\xi+1)(\alpha-3)}{\vartheta_i(\alpha-2)(\xi+1)}\right)g_{\alpha,\vartheta_i}\left(\xi+1\right)\\ &-\left(\frac{\vartheta_i(\alpha-1)^2-\xi (\alpha-3)}{\vartheta_i(\alpha-2)\xi}\right)
  g_{\alpha,\vartheta_i}\left(\xi\right)+\frac{G_{\alpha,\vartheta_i}(1+\xi)-G_{\alpha,\vartheta_i}(\xi)}{(\alpha-2)\vartheta_i^2}+\frac{g_{\alpha,\vartheta_i}(1+\xi)^2}{1-  G_{\alpha,\vartheta_i}(1+\xi)}.
\end{split}
\end{equation*}

\subsection{Fisher Information Matrix for the Two-tiered Gamma Model}\label{app2}
First, with \eqref{expcengam} and \eqref{expcengam2}, we get
\begin{equation*}
  \begin{split}
E_{\theta}\left[ \frac{\partial \ell_i}{\partial \beta_k}\frac{\partial \ell_i}{\partial \beta_l}
   \right]=E_{\theta}&\left[x_{ik}x_{il}\left(\frac{y_i+\xi}{\vartheta_i}-a-\frac{\xi\cdot
    g_{\alpha,\vartheta_i}\left(\xi\right)}{\vartheta_i\cdot(1-
    G_{\alpha,\vartheta_i}(\xi))}\right)^2\mathbf{1}_{\{0<  y_i <  1\}} \right]\\
+&E_{\theta}\left[\frac{x_{ik}x_{il}}{\vartheta_i^2}\cdot\left(\frac{(1+\xi)\cdot
      g_{\alpha,\vartheta_i}\left(1+\xi\right)}{1- G_{\alpha,\vartheta_i}(1+\xi)}-\frac{\xi\cdot
      g_{\alpha,\vartheta_i}\left(\xi\right)}{1-
      G_{\alpha,\vartheta_i}(\xi)}\right)^2\mathbf{1}_{\{ y_i =  1\}}\right]\\
=x_{ik}&x_{il}a\frac{(G_{\alpha+2,\vartheta_i}(1+\xi)-G_{\alpha+2,\vartheta_i}(\xi))(1- G_{\tilde{\vartheta}_i,a}(\xi))}{1-
  G_{\alpha,\vartheta_i}(\xi)}\\
+&x_{ik}x_{il}\frac{\alpha^2\left(\xi g_{\alpha+1,\vartheta_i}\left(\xi\right)-(1+\xi)g_{\alpha+1,1}\left(\frac{1+\xi}{\vartheta_i}\right)\right)(1- G_{\tilde{\vartheta}_i,a}(\xi))}{(\alpha+1)(1- G_{\alpha,\vartheta_i}(\xi))}\\
+&x_{ik}x_{il}\frac{\alpha\left((1+\xi)g_{\alpha,\vartheta_i}\left(1+\xi\right)-\xi g_{\alpha,\vartheta_i}\left(\xi\right)\right)(1- G_{\tilde{\vartheta}_i,a}(\xi))}{1- G_{\alpha,\vartheta_i}(\xi)}\\-&x_{ik}x_{il}\frac{\xi^2g_{\alpha,\vartheta_i}\left(\xi\right)^2(1- G_{\tilde{\vartheta}_i,a}(\xi))}{(1- G_{\alpha,\vartheta_i}(\xi))^2}\\+&x_{ik}x_{il}\frac{(1+\xi)^2g_{\alpha,\vartheta_i}\left(1+\xi\right)^2(1- G_{\tilde{\vartheta}_i,a}(\xi))}{(1- G_{\alpha,\vartheta_i}(\xi))(1- G_{\alpha,\vartheta_i}(1+\xi))}
\end{split}
\end{equation*}
Next, with \eqref{expcengam} and the identity in \eqref{gammaidentity3}, we get
\begin{equation*}
  \begin{split}
E_{\theta}\left[ \frac{\partial \ell_i}{\partial \beta_k}\frac{\partial \ell_i}{\partial \gamma_l}
   \right]=E_{\theta}&\left[x_{ik}x_{il}\left(\frac{y_i+\xi}{\vartheta_i}-a-\frac{\xi\cdot
    g_{\alpha,\vartheta_i}\left(\xi\right)}{1-
    G_{\alpha,\vartheta_i}(\xi)}\right)\frac{\xi\cdot  g_{\alpha,\tilde{\vartheta}_i}\left(\xi\right)}{1- G_{\tilde{\vartheta}_i,a}(\xi)}\mathbf{1}_{\{0<  y_i <  1\}} \right]\\
+E_{\theta}&\left[x_{ik}x_{il}\cdot\left(\frac{(1+\xi)\cdot
      g_{\alpha,\vartheta_i}\left(1+\xi\right)}{1- G_{\alpha,\vartheta_i}(1+\xi)}-\frac{\xi\cdot
      g_{\alpha,\vartheta_i}\left(\xi\right)}{1-
      G_{\alpha,\vartheta_i}(\xi)}\right)\frac{\xi\cdot  g_{\alpha,\tilde{\vartheta}_i}\left(\xi\right)}{1- G_{\tilde{\vartheta}_i,a}(\xi)}\mathbf{1}_{\{ y_i =  1\}}\right]\\
= & 0.
\end{split}
\end{equation*}
Finally, we calculate
\begin{equation*}
  \begin{split}
E_{\theta}\left[ \frac{\partial \ell_i}{\partial \gamma_k}\frac{\partial \ell_i}{\partial \gamma_l}
   \right]=E_{\theta}&\left[x_{ik}x_{il}\left(\frac{\xi\cdot  g_{\alpha,\tilde{\vartheta}_i}\left(\xi\right)}{\tilde{\vartheta}_i\cdot G_{\tilde{\vartheta}_i,a}(\xi)}\right)^2\mathbf{1}_{\{ y_i =  0\}} \right]\\
+&E_{\theta}\left[x_{ik}x_{il}\left(\frac{\xi\cdot  g_{\alpha,\tilde{\vartheta}_i}\left(\xi\right)}{\tilde{\vartheta}_i\cdot(1- G_{\tilde{\vartheta}_i,a}(\xi))}\right)^2\left(\mathbf{1}_{\{0<  y_i <  1\}}+\mathbf{1}_{\{ y_i =  1\}}\right) \right]\\
= & x_{ik}x_{il}\frac{\xi^2\cdot
  g_{\alpha,\tilde{\vartheta}_i}\left(\xi\right)^2}{
  G_{\tilde{\vartheta}_i,a}(\xi)}+x_{ik}x_{il}\frac{\xi^2\cdot
  g_{\alpha,\tilde{\vartheta}_i}\left(\xi\right)^2}{1-
  G_{\tilde{\vartheta}_i,a}(\xi)}\\
=&x_{ik}x_{il}\frac{\xi^2\cdot
  g_{\alpha,\tilde{\vartheta}_i}\left(\xi\right)^2}{ G_{\tilde{\vartheta}_i,a}(\xi)(1-
  G_{\tilde{\vartheta}_i,a}(\xi))}.
\end{split}
\end{equation*}

\subsection{Useful Identities and Integrals}\label{IdInt}
By partial integration, we calculate
\begin{align*}
G_{\alpha+1,\vartheta}(\xi)=&\frac{1}{\vartheta^{\alpha+1}\Gamma(\alpha+1)}\int_0^h{y^{\alpha}\exp(-y/\vartheta)dy}\\
=&\frac{1}{\vartheta^{\alpha+1}\Gamma(\alpha+1)}\left(-\xi^{\alpha}s\exp(-\xi
  /\vartheta)\right)\\ 
&+\frac{1}{\vartheta^{\alpha+1}\Gamma(\alpha+1)}\int_0^h{ay^{\alpha-1}s\exp(-y/\vartheta)dy}\\
=&-\frac{1}{\Gamma(\alpha+1)}\left(\frac{\xi}{\vartheta}\right)^{\alpha}\exp(-\xi /\vartheta)+\frac{1}{\vartheta^{\alpha}\Gamma(\alpha)}\int_0^h{y^{\alpha-1}\exp(-y/\vartheta)dy}\\
=&-\vartheta g_{\alpha+1,\vartheta}\left(\xi\right)+G_{\alpha,\vartheta}(\xi).
\end{align*}

And from this follows
\begin{equation}\label{gammaidentity2}
G_{\alpha+1,\vartheta}(\xi)-G_{\alpha,\vartheta}(\xi)=-\vartheta g_{\alpha+1,\vartheta}\left(\xi\right)
\end{equation}
or
\begin{equation}\label{gammaidentity3}
G_{\alpha+1,\vartheta}(\xi)-G_{\alpha,\vartheta}(\xi)=-\frac{\xi}{\alpha}g_{\alpha,\vartheta}\left(\xi\right).
\end{equation}

For $0\leq l<u$, the following equations hold true.
\begin{align}
\int_l^u{yg_{\alpha,\vartheta}(y)dy}&=\alpha\vartheta(G_{\alpha+1,\vartheta}(u)-G_{\alpha+1,\vartheta}(l)).\label{expcengam}
\end{align}

\begin{align}
\int_l^u{y^2g_{\alpha,\vartheta}(y)dy}&=\vartheta^2a(\alpha+1)(G_{\alpha+2,\vartheta}(u)-G_{\alpha+2,\vartheta}(l)).\label{expcengam2}
\end{align}

\begin{align}
\int_l^u{\frac{1}{y}g_{\alpha,\vartheta}(y)dy}&=\frac{1}{(\alpha-1)\vartheta }(G_{\alpha-1,\vartheta}(u)-G_{\alpha-1,\vartheta}(l)).\label{expcengam-1}
\end{align}

\begin{align}
\int_l^u{\frac{1}{y^2}g_{\alpha,\vartheta}(y)dy}&=\frac{1}{(\alpha-1)(\alpha-2)\vartheta^2}(G_{\alpha-2,\vartheta}(u)-G_{\alpha-2,\vartheta}(l)).\label{expcengam-2}
\end{align}

\begin{align}
\int_l^u{\log(y)g_{\alpha,\vartheta}(y)dy}&=\log(\vartheta)(G_{\alpha,\vartheta}(u)-G_{\alpha,\vartheta}(l))+H^{(1)}_{\alpha}\left(\frac{l}{\vartheta},\frac{u}{\vartheta}\right).\label{expcengamlog}
\end{align}

\begin{align}
\int_l^u{\log(y)^2g_{\alpha,\vartheta}(y)dy}=&\log(\vartheta)^2(G_{\alpha,\vartheta}(u)-G_{\alpha,\vartheta}(l))\nonumber\\ &+2\log(\vartheta)H^{(1)}_{\alpha}\left(\frac{l}{\vartheta},\frac{u}{\vartheta}\right)+H^{(2)}_{\alpha}\left(\frac{l}{\vartheta},\frac{u}{\vartheta}\right).\label{expcengamlog2}
\end{align}

\begin{align} 
\int_l^u{y\log(y)g_{\alpha,\vartheta}(y)dy}=&\alpha\vartheta\log(\vartheta)(G_{\alpha+1,\vartheta}(u)-G_{\alpha+1,\vartheta}(l))\nonumber\\
&+\alpha\vartheta H_{\alpha+1}^{(1)}\left(\frac{l}{\vartheta},\frac{u}{\vartheta}\right).\label{expcengamgamlog}
\end{align}

\begin{align}
\int_l^u{\frac{\log(y)}{y}g_{\alpha,\vartheta}(y)dy}=&\frac{1}{(\alpha-1)\vartheta }\log(\vartheta)(G_{\alpha-1,\vartheta}(u)-G_{\alpha-1,\vartheta}(l))\nonumber\\ &+\frac{1}{(\alpha-1)\vartheta }H_{\alpha-1}^{(1)}\left(\frac{l}{\vartheta},\frac{u}{\vartheta}\right).\label{expcengam-1gamlog}
\end{align}
\subsection{Descriptive statistics for covariates}\label{AppendixDesc}
\begin{table}[ht]
\begin{center}
\begin{tabular}{l|rrrr}
\hline \hline
&Mean&Standard Deviation\\\hline
RDT&0.65 & 0.30  \\
Face Value (log) & 3.84 &0.54\\
Insured  Fraction & 0.06&0.03\\
\hline
&Low / Small & Mid / Medium & High / Large\\\hline
Experience & 15.52&55.38&29.10\\
Size & 85.06&9.95&4.98\\
 \hline
&Maintenance & Hybrid & Performance & Other\\\hline
Type & 88.05 & 6.85&4.42&0.67\\
\hline
\end{tabular}
\end{center}
\caption{Descriptive statistics for covariates. For categorical variables,
  the frequency (in $\%$) of the levels are given.}
\label{tab:SumStat}
\end{table}
\subsection{Additional Plots Illustrating Other Fitted
  Models}\label{Appendix2}
Additionally, two different types of models have been fitted two the data. First, the
two-limit version of the normal Tobit model and its corresponding
two-tiered and zero-inflated extensions. Further, we fitted models using the skewed
t-distribution (\citet{AzCa03}) where, in each model, the shifted Gamma
distribution is replaced by a skewed t-distribution. The degrees of freedom
were chosen to be $1$ since this provided the best fit in general.

\begin{figure}[h]
\centering
\includegraphics[scale=0.4]{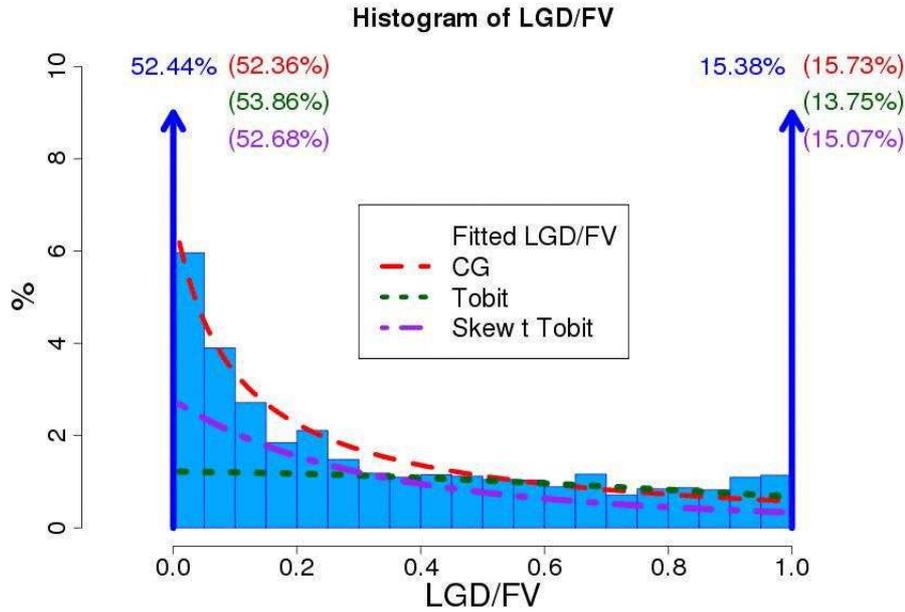} 
\caption{Comparison of fitted censored gamma, normal Tobit, and skew t Tobit models with no
  covariates. The numbers above the blue arrows represent the
  percentage of LGD/FV's being exactly zero or one, respectively. In
  parentheses are the corresponding numbers as predicted by the  models.} 
\label{fig:TobitnoCov}
\end{figure}

\begin{figure}[h]
\centering
\includegraphics[scale=0.4,angle=0]{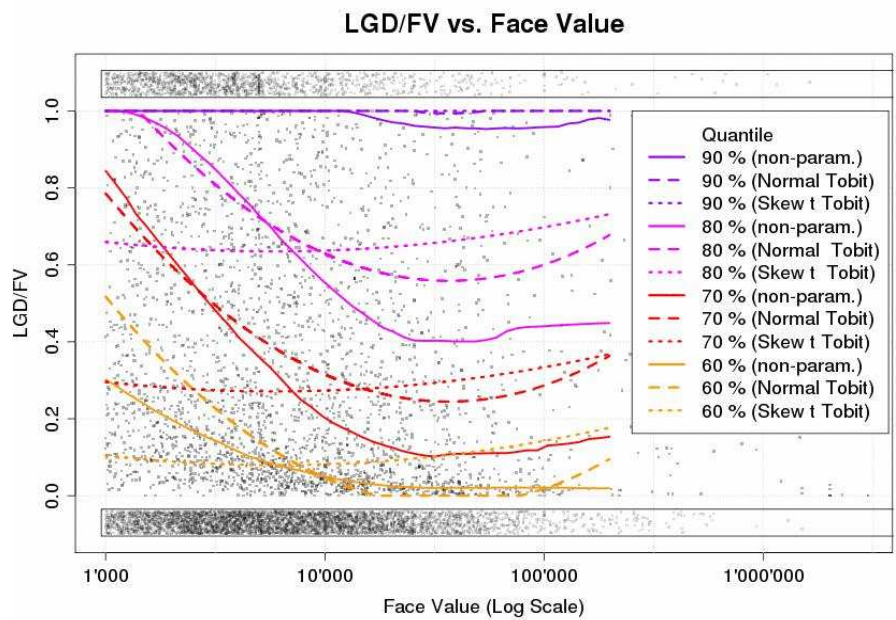} 
\caption{Scatter plot of LGD/FV versus face value (on a logarithmic scale). The jittered points in the bars below $0.0$ and above $1.0$ represent bonds with
  LGD/FV being exactly zero and one, respectively. The colored solid lines
  are non-parametrically fitted quantiles and the mean. The dashed and dotted lines
  represent quantiles of the fitted normal Tobit model and the skew t Tobit
  model, respectively. Logarithmic and squared logarithmic
  face value are taken as covariates.}
\label{fig:TobitFVCov}
\end{figure}

\begin{sidewaystable}[ht]
\begin{center}
\begin{tabular}{ll|rl|rlrl|rlrl}
\hline\hline
&  \multicolumn{1}{r|}{Model} &\multicolumn{2}{c}{Censored}&\multicolumn{4}{|c|}{Two-Tiered}&\multicolumn{4}{c}{Zero-Inflated}\\ 
Covariate & &  Coef & Std. Err. & Coef($\boldsymbol{\beta}$) & Std. Err. &Coef($\boldsymbol{\gamma}$) & Std. Err. & Coef($\boldsymbol{\beta}$) & Std. Err. &Coef($\boldsymbol{\gamma}$) & Std. Err. \\
\hline
 Intercept & & 0.86 & 0.13 *** & 15. & 0.030 *** & 15. & 0.030 *** & 2.1 & 0.19 *** & 2.1 & 0.19 *** \\
 \hline \multirow{2}{*}{RDT} &\text{Lin} & -0.19 & 0.044 *** & 1.5 & 0.049 *** & 0.41 & 0.33   & -0.042 & 0.071   & 4.5 & 1.1 *** \\
   &\text{Quad} & -0.27 & 0.15 $\cdot$ & 8.3 & 0.087 *** & -0.70 & 0.11 *** & 0.33 & 0.24   & 0.59 & 0.20 ** \\
 \hline \multirow{2}{*}{Experience} &\text{Lin} & -0.40 & 0.028 *** & -2.0 & 0.0041 *** & -1.6 & 0.38 *** & -0.32 & 0.043 *** & 2.3 & 0.65 *** \\
   &\text{Quad} & 0.036 & 0.020 $\cdot$ & 0.29 & 0.0091 *** & -0.87 & 0.073 *** & 0.013 & 0.027   & 0.36 & 0.17 * \\
 \hline \multirow{2}{*}{Size} &\text{Lin} & 0.40 & 0.16 * & 2.8 & 0.051 *** & 0.036 & 0.053   & 0.23 & 0.21   & -0.094 & 0.092   \\
   &\text{Quad} & 0.51 & 0.10 *** & 0.10 & 0.0098 *** & 0.91 & 0.40 * & 0.049 & 0.13   & -0.27 & 0.51   \\
 \hline \multirow{2}{*}{Face Value} &\text{Lin} & -0.22 & 0.026 *** & -5.4 & 0.0017 *** & 1.4 & 0.25 *** & -0.50 & 0.042 *** & -1.5 & 0.31 *** \\
   &\text{Quad} & 0.22 & 0.025 *** & 0.27 & 0.0021 *** & -0.047 & 0.068   & 0.26 & 0.034 *** & -1.5 & 0.31 *** \\
 \hline \multirow{3}{*}{Type} & Hybrid & 1.4 & 0.47 ** & 8.9 & 0.021 *** & 0.64 & 0.075 *** & 1.0 & 0.46 * & -0.78 & 0.32 * \\
   & Performance & -0.036 & 0.052   & 0.96 & 1.7   & 3.3 & 1.5 * & 0.0095 & 0.061   & -3.5 & 2.7   \\
   & Other & -0.019 & 0.071   & 2.4 & 0.031 *** & -0.16 & 0.14   & 0.035 & 0.10   & 0.19 & 0.31   \\
 \hline Ins. Frac. & & 0.29 & 0.19   & 8.7 & 0.99 *** & -0.28 & 0.19   & 0.76 & 0.22 *** & 0.26 & 0.37   \\
  \hline &  & Value & Std. Err.   &  & Value & Std. Err. &  &  &Value &  Std. Err.& \\
 \hline $\log(\sigma)$ &  & -0.040 & 0.016   & & 0.81 & 0.0014   & & & -0.15 & 0.022   & \\
\hline
\multicolumn{2}{c|}{Log-Likelihood}  &  \multicolumn{2}{c}{ -8241.6 } &\multicolumn{4}{|c|}{ -7864.4 } &\multicolumn{4}{c}{ -8169 } \\ 
\hline
\multicolumn{2}{c|}{AIC}  &  \multicolumn{2}{c}{ 16511.2 } &\multicolumn{4}{|c|}{ 15780.8 } &\multicolumn{4}{c}{ 16390 } \\ 
\hline
\end{tabular}
\end{center}
\caption{Fitted censored, two-tiered, and zero-inflated normal Tobit models including all covariates. Codes for significance levels:   '***': $p<0.001$,  '**': $0.001\leq p < 0.01$,  '*':  $0.01 \leq p < 0.05$, '.': $0.05 \leq p < 0.1$.}
\label{tab:fitTobitfull}
\end{sidewaystable}

\begin{sidewaystable}[ht]
\begin{center}
\begin{tabular}{ll|rl|rlrl|rlrl}
\hline\hline
&  \multicolumn{1}{r|}{Model} &\multicolumn{2}{c}{Censored}&\multicolumn{4}{|c|}{Two-Tiered}&\multicolumn{4}{c}{Zero-Inflated}\\ 
Covariate & &  Coef & Std. Err. & Coef($\boldsymbol{\beta}$) & Std. Err. &Coef($\boldsymbol{\gamma}$) & Std. Err. & Coef($\boldsymbol{\beta}$) & Std. Err. &Coef($\boldsymbol{\gamma}$) & Std. Err. \\
\hline
 Intercept & & -0.38 & 0.056   & -1.9 & 0.10   & 0.015 & 0.0090 $\cdot$ & -0.090 & 0.075   & -0.49 & 0.33   \\
 \hline \multirow{2}{*}{RDT} &\text{Lin} & -0.15 & 0.023 *** & -0.15 & 0.037 *** & -0.010 & 0.0025 *** & -0.19 & 0.041 *** & -0.17 & 0.16   \\
   &\text{Quad} & -0.38 & 0.073 *** & -0.36 & 0.12 ** & -0.026 & 0.0080 ** & -0.39 & 0.14 ** & 0.23 & 0.54   \\
 \hline \multirow{2}{*}{Experience} &\text{Lin} & -0.12 & 0.013 *** & 0.20 & 0.034 *** & -0.016 & 0.0037 *** & 0.050 & 0.026 $\cdot$ & 1.3 & 0.21 *** \\
   &\text{Quad} & -0.013 & 0.0094   & -0.060 & 0.022 ** & 0.0020 & 0.0011 $\cdot$ & -0.0017 & 0.023   & -0.28 & 0.15 $\cdot$ \\
 \hline \multirow{2}{*}{Size} &\text{Lin} & 0.26 & 0.10 ** & -0.20 & 0.13   & 0.012 & 0.0060 $\cdot$ & 0.42 & 0.15 ** & 0.48 & 0.41   \\
   &\text{Quad} & 0.33 & 0.069 *** & 0.0064 & 0.073   & 0.021 & 0.0048 *** & 0.33 & 0.10 ** & -0.11 & 0.30   \\
 \hline \multirow{2}{*}{Face Value} &\text{Lin} & 0.038 & 0.010 *** & 0.40 & 0.019 *** & 0.0024 & 0.00079 ** & 0.030 & 0.0074 *** & 0.040 & 0.066   \\
   &\text{Quad} & 0.042 & 0.0064 *** & -0.12 & 0.011 *** & 0.011 & 0.0028 *** & 0.0069 & 0.016   & -0.44 & 0.080 *** \\
 \hline \multirow{3}{*}{Type} & Hybrid & 0.56 & 0.11 *** & -2.1 & 0.48 *** & 0.066 & 0.033 * & 0.020 & 0.22   & -1.8 & 1.1 $\cdot$ \\
   & Performance & -0.038 & 0.021 $\cdot$ & -0.0033 & 0.026   & -0.0038 & 0.0024   & -0.034 & 0.037   & -0.038 & 0.19   \\
   & Other & -0.12 & 0.0099 *** & -0.050 & 0.031   & -0.0070 & 0.0033 * & -0.00075 & 0.062   & 0.25 & 0.21   \\
 \hline Ins. Frac. & & -0.13 & 0.085   & -1.1 & 0.64 $\cdot$ & -0.013 & 0.0098   & 0.50 & 0.084 *** & 1.1 & 0.34 ** \\
  \hline &  & Value & Std. Err.   &  & Value & Std. Err. &  &  &Value &  Std. Err.& \\
 \hline \multirow{3}{*}{Skew t Par.} & $\nu$ & 1 & & & 1 & & & & 1 & \\
   &$\log(\sigma)$ & -1.1 & 0.028   & & -3.5 & 0.22   & & & -0.90 & 0.033   & \\
   & $\alpha$ & 30. & 23.   & & -1.0 & 0.31   & & & 38. & 27.   & \\
\hline
\multicolumn{2}{c|}{Log-Likelihood}  &  \multicolumn{2}{c}{ -8019 } &\multicolumn{4}{|c|}{ -7692.4 } &\multicolumn{4}{c}{ -7964.4 } \\ 
\hline
\multicolumn{2}{c|}{AIC}  &  \multicolumn{2}{c}{ 16067.9 } &\multicolumn{4}{|c|}{ 15440.7 } &\multicolumn{4}{c}{ 15984.8 } \\ 
\hline
\end{tabular}
\end{center}
\caption{Fitted censored, two-tiered, and zero-inflated skew t (df =1) Tobit models including all covariates. Codes for significance levels:   '***': $p<0.001$,  '**': $0.001\leq p < 0.01$,  '*':  $0.01 \leq p < 0.05$, '.': $0.05 \leq p < 0.1$.}
\label{tab:fitSkewTfull}
\end{sidewaystable}

\end{document}